\documentclass[preprintnumbers,prc]{revtex4-1}

\usepackage{graphicx}
\usepackage{amssymb}
\usepackage{epstopdf}
\usepackage{dcolumn}
\usepackage{bm}
\begin{document}
    
\title{Measurement of the neutron decay electron-antineutrino angular correlation by the aCORN experiment}
    
\author{M.~T.~Hassan}\thanks{current address: Los Alamos National Laboratory, Los Alamos, NM 87545, USA.}
\affiliation{Department of Physics and Engineering Physics, Tulane University, New Orleans, LA 70118}
\author{W.~A.~Byron}\thanks{current address: Dept. of Physics, University of Washington, Seattle, WA 98195, USA}
\affiliation{Department of Physics and Engineering Physics, Tulane University, New Orleans, LA 70118}
\author{G.~Darius}
\affiliation{Department of Physics and Engineering Physics, Tulane University, New Orleans, LA 70118}
\author{C.~DeAngelis}
\affiliation{Department of Physics and Engineering Physics, Tulane University, New Orleans, LA 70118}
\author{F.~E.~Wietfeldt}
\affiliation{Department of Physics and Engineering Physics, Tulane University, New Orleans, LA 70118}
\author{B.~Collett}
\affiliation{Physics Department, Hamilton College, Clinton, NY 13323}
\author{G.~L.~Jones}
\affiliation{Physics Department, Hamilton College, Clinton, NY 13323}
\author{A.~Komives}
\affiliation{Department of Physics and Astronomy, DePauw University, Greencastle, IN 46135}
\author{G.~Noid}
\affiliation{CEEM, Indiana University, Bloomington, IN 47408}
\author{E.~J.~Stephenson}
\affiliation{CEEM, Indiana University, Bloomington, IN 47408}
\author{F.~Bateman}
\affiliation{National Institute of Standards and Technology, Gaithersburg, MD 20899, USA}
\author{M.~S.~Dewey}
\affiliation{National Institute of Standards and Technology, Gaithersburg, MD 20899, USA}
\author{T.~R.~Gentile}
\affiliation{National Institute of Standards and Technology, Gaithersburg, MD 20899, USA}
\author{M.~P.~Mendenhall}\thanks{current address: Lawrence Livermore National Laboratory, Livermore, CA 94550, USA}
\affiliation{National Institute of Standards and Technology, Gaithersburg, MD 20899, USA}
\author{J.~S.~Nico}
\affiliation{National Institute of Standards and Technology, Gaithersburg, MD 20899, USA}

\date{\today}

\begin{abstract}
The aCORN experiment measures the neutron decay electron-antineutrino correlation ($a$-coefficient) using a novel method based on an asymmetry in proton time-of-flight for events where the beta electron and recoil proton are detected in delayed coincidence. We report the data analysis and result from the second run at the NIST Center for Neutron Research, using the high-flux cold neutron beam on the new NG-C neutron guide end position: $a = -0.10758 \pm 0.00136 (\mbox{stat}) \pm 0.00148 (\mbox{sys})$. This is consistent within uncertainties with the result from the first aCORN run on the NG-6 cold neutron beam. Combining the two aCORN runs we obtain $a = -0.10782 \pm 0.00124 (\mbox{stat}) \pm 0.00133 (\mbox{sys})$, which has an overall relative standard uncertainty of 1.7 \%. The corresponding result for the ratio of weak coupling constants $\lambda = G_A/G_V$ is $\lambda = -1.2796\pm 0.0062$.
\end{abstract}

\maketitle

\section{Introduction}
\label{S:Intro}
The free neutron decays into a proton, electron, and antineutrino via the charged-current weak interaction. This is the simplest example of nuclear beta decay. In contrast to beta decay of most nuclei, the dynamics of neutron decay are undisturbed by nuclear structure effects. Experimental observables can be directly related to fundamental parameters in the theory.  As a result, neutron decay is an excellent laboratory for studying details of the weak nuclear force and searching for hints of physics beyond the Standard Model (SM). 
The important experimental features of neutron decay are described by the formula of Jackson, Treiman, and
Wyld \cite{JTW}, which gives the decay probability $N$  of a spin-1/2 beta decay system in terms of the neutron spin polarization $\mbox{\boldmath $P$}$, the beta electron total energy and momentum $E_e$, $\mbox{\boldmath $p_e$}$, and the antineutrino total energy and momentum $E_{\nu}$, $\mbox{\boldmath $p_{\nu}$}$
\begin{equation}
\label{E:JTWeqn}
N \propto \frac{1}{\tau_n}E_e |{\mbox{\boldmath $p_e$}}| (Q-E_e)^2 \left[ 1 
+ a\frac{\mbox{\boldmath $p_e$}\cdot \mbox{\boldmath $p_\nu$}}{E_e E_\nu} + b\frac{m_e}{E_e}
+ \mbox{\boldmath $P$}\cdot \left( A\frac{\mbox{\boldmath $p_e$}}{E_e} + B\frac{\mbox{\boldmath $p_\nu$}}{E_\nu}
+ D\frac{({\mbox{\boldmath $p_e$}}\times \mbox{\boldmath $p_\nu$})}{E_e E_\nu} \right) \right].
\end{equation}

$Q$ = 1293 keV is the neutron-proton mass difference, $m_e$ is the electron mass, and $\tau_n$ is the neutron lifetime. Here and throughout velocity is in units with $c=1$. The parameters $a$, $A$, $B$, and $D$ are correlation coefficients which are measured by experiment. We note that $a$, $b$ are parity conserving, $A$, $B$ are parity violating, and $D$ violates time-reversal symmetry. The Fierz interference parameter $b$ is zero in the SM; it would be generated by the presence of scalar or tensor weak currents. Neglecting recoil order effects,  the values of the other coefficients are related to two basic parameters in the theory: the nucleon weak vector and axial vector coupling constants $G_V$ and $G_A$. Writing their ratio as $\lambda = G_A/G_V$ we have  \cite{JTW}
\[
\tau_n = \frac{2 \pi^3 \hbar^7}{(G_V^2 + 3G_A^2) m_e^5\, f_R} \qquad 
a  =  \frac{1 - \lambda^2}{1 + 3\lambda^2} 
\]
\begin{equation}
\label{E:SMtaABD}
A  =  -2\frac{{\rm Re}\{\lambda \} + \lambda^2}{1 + 3\lambda^2} \qquad
B  =  -2\frac{{\rm Re}\{\lambda \} - \lambda^2}{1 + 3\lambda^2} \qquad
D  =  2\frac{{\rm Im}\{\lambda \}}{1 + 3\lambda^2}
\end{equation}
where $f_R$ is the value of the integral over the Fermi energy spectrum. There are two main motivations for precision measurements of neutron decay observables.
\par
The first is to accurately determine the values of $G_V$ and $G_A$. These constants appear not only in neutron decay but in many other weak interaction processes involving free neutrons and protons that are important in astrophysics, cosmology, solar physics, and neutrino detection \cite{Dub91,Bar05}. The value of $G_V$ gives the first element $V_{ud}$ of the Cabibbo-Kobayashi-Maskawa (CKM) quark mixing matrix: $G_V = G_FV_{ud}$, where $G_F$ is the universal weak coupling constant obtained from the muon lifetime. A very important low energy test of the Standard Model is the unitarity of the first row of the CKM matrix
\begin{equation}
\label{E:CKMu}
|V_{ud}|^2 + |V_{us}|^2 + |V_{ub}|^2 = 1.
\end{equation}
The term $|V_{ub}|^2$ is small enough to be neglected so in practice this is a precise comparison of $V_{ud}$ and $V_{us}$. A real violation of this unitarity condition would be a clear sign of new physics Beyond the Standard Model (BSM) at the low energy, precision frontier. For example supersymmetry loop corrections could cause a departure from equation \ref{E:CKMu} at the few $10^{-4}$ level and reveal new physics that lies beyond present constraints from the Large Hadron Collider \cite{Bau13}.
\par
The second motivation is to search for small discrepancies in the values of these observables that could result from BSM physics. We see from equation \ref{E:SMtaABD} that a measurement of $\tau_n$ and any one of $a$, $A$, or $B$ determine the real values of  $G_V$ and $G_A$, but new physics could introduce dependencies on additional new parameters. A useful model-independent self-consistency test is obtained from the Mostovoy parameters \cite{Mo76}
\begin{eqnarray}
\label{E:F1F2}
F_1  & = & 1 + A - B - a = 0 \nonumber\\
F_2  & = & aB - A - A^2 = 0.
\end{eqnarray}
which follow algebraically from the relations in equation \ref{E:SMtaABD}. Inserting the Particle Data Group 2020 (PDG 2020) \cite{PDG20} recommended values we have $F_1 = 0.0056 \pm 0.0041$ and 
$F_2 = 0.0014 \pm 0.0028$, consistent with the SM expectations. The PDG 2020 experimental uncertainty in the $a$-coefficient is the largest contributor to the uncertainties in $F_1$ and $F_2$.
We note that recoil order corrections will cause $F_1$ and $F_2$ to differ from zero at the 10$^{-4}$ level, but those corrections are calculable. Important model-dependent tests for new physics can be made with neutron decay observables. The relative values of $a$, $A$, and $B$ can be related to the strength of hypothetical right-handed weak forces and scalar and tensor forces \cite{Iv13,Bh16}. Gardner and Zhang have shown that a comparison of $a$ and $A$ at the $10^{-3}$ level can place sharp limits on possible conserved-vector-current (CVC) violation and second-class currents \cite{Gar01}. Possible extensions to the Standard Model, such as supersymmetry or left-right symmetric models, could lead to observable departures from the predictions in equations \ref{E:SMtaABD}. 
\par
\begin{figure}
\centering
\includegraphics[width = 6.5in]{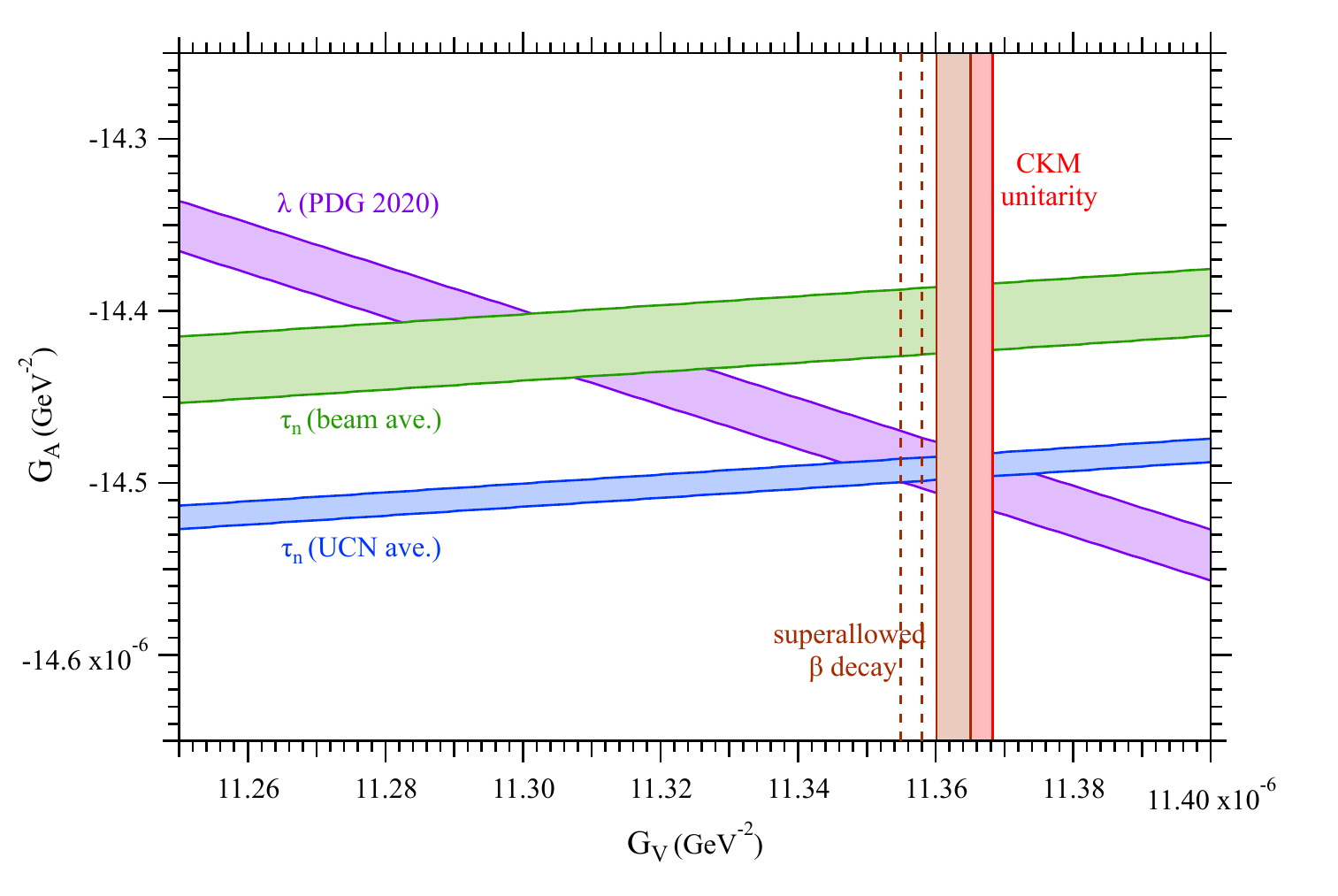}
\vspace{-0.2in}
\caption{\label{F:gAgV} A summary of experimental constraints on the nucleon weak coupling constants $G_A$ and $G_V$. The purple band is the PDG 2020 \cite{PDG20} recommended value for $\lambda$ from neutron decay parameters $A$ and $a$, including a scale factor of $\sqrt{\chi_\nu^2} = 2.6$ to account for poor agreement among experiments. The green (no scale factor) and blue (scale factor $\sqrt{\chi_\nu^2} = 1.5$) bands are derived from the neutron lifetime averages for the beam and UCN storage experiments. The brown vertical band shows $G_V$ from superallowed beta decay \cite{Har15} and the dashed lines indicate the shift due to the calculation of $\Delta_R$ by Seng, {\em et al.} \cite{Sen18}. The red vertical band shows the CKM matrix unitarity condition using the PDG recommended value of $V_{us}$ \cite{PDG20}.}
\end{figure}

Figure \ref{F:gAgV} summarizes the current experimental results for $G_A$ and $G_V$. The PDG 2020 recommended value $\lambda=-1.2756 \pm 0.0013$ includes nine measurements of the neutron decay coefficients $A$ and $a$ from 1986 to 2019, and the uncertainty is expanded by a factor of 2.6 due to poor agreement. The most recent and precise results for the beta asymmetry $A$ from the PERKEO II,III \cite{Mun13,Mar19} and UCNA \cite{Bro18} experiments are in good agreement and give a more negative value of $\lambda$. The neutron lifetime averages from beam method and ultracold storage experiments significantly disagree, see for example \cite{Wie11}. The value of $G_V = 1.13625(24) \times 10^{-5}$ GeV$^{-2}$ from an evaluation of 222 measurements of 20 superallowed beta decay systems \cite{Har15} agrees moderately with the CKM unitarity requirement (using the PDG 2020 $V_{us}$ \cite{PDG20}), differing by 1.2$\sigma$. But a 2018 calculation of electroweak box diagram contributions to the ``universal'' radiative correction $\Delta_R$ by Seng, {\em et al.} \cite{Sen18} shifts the superallowed result down to $G_V = 1.13570(16) \times 10^{-5}$ GeV$^{-2}$ which would violate CKM unitarity by 3.4$\sigma$. In a recent paper Czarnecki, Marciano, and Sirlin \cite{Car19} recommend an intermediate value for $\Delta_R$ and hence $G_V$. Clearly the present experimental situation with $G_A$ and $G_V$ is not satisfactory and additional measurements, with careful attention to evaluating systematic effects and uncertainties, are needed.

\section{Experimental method}
\label{S:aMethod}
\begin{figure}
\centering
\includegraphics[width = 6.5in]{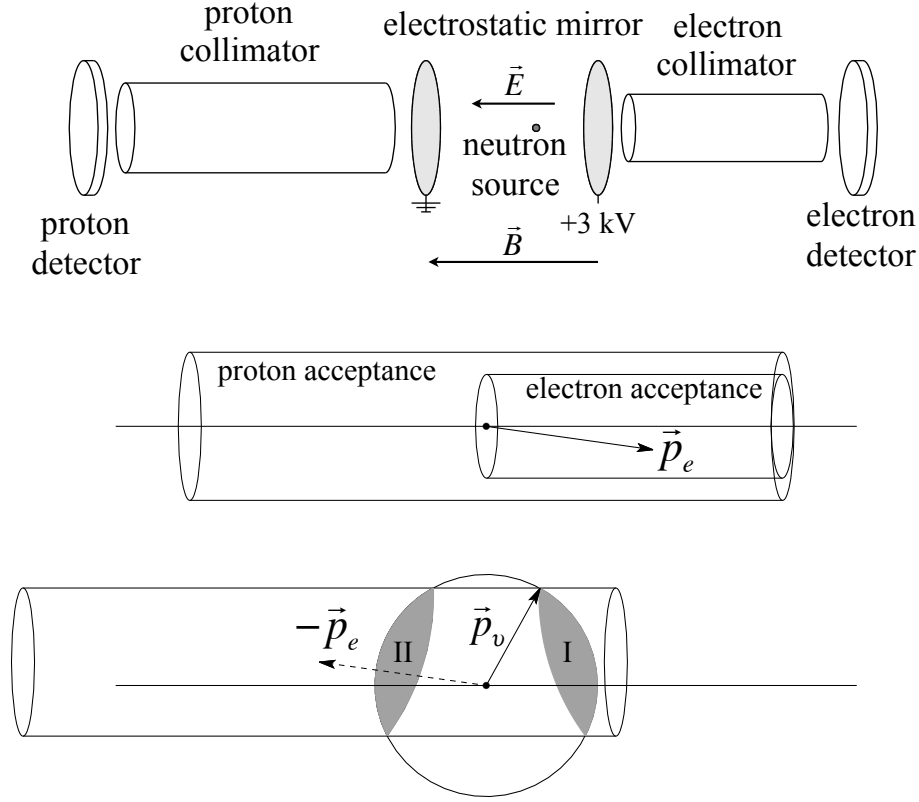}
\vspace{-0.2in}
\caption{\label{F:cylfig} An illustration of the aCORN experimental method. Top: A neutron source, shown as a point source here, lies on axis between a set of proton and electron detectors. A uniform axial magnetic field $\vec{B}$ is present throughout.  Electron and proton collimators act to limit the transverse momenta of detected electrons and protons from neutron decay. An electrostatic mirror produces an approximately uniform electric field $\vec{E}$ in the decay region that accelerates and directs all protons toward the proton detector, but beta electrons in the energy range of interest must be emitted into the right hemisphere to be detected. Middle: A momentum space plot showing the cylindrical momentum acceptances of electrons and protons. Bottom: A momentum space construction of the acceptance for antineutrinos from neutron decay, when the detected electron momentum was $\vec{p_e}$ as shown and the proton was also detected. Conservation of energy and momentum restricts the antineutrino momentum to the shaded regions I and II which have equal solid angle from the source. Region I is correlated with $\vec{p_e}$ and region II is anticorrelated, so the asymmetry in events associated with each region measures the $a$-coefficient.}
\end{figure}

The traditional method for measuring the $a$-coefficient is from the shape of the recoil ion energy spectrum. If the beta electron and antineutrino momenta are anticorrelated, the average recoil momentum is reduced, which shifts weight to the low-energy part of the spectrum. Until recently all measurements of the neutron $a$-coefficient used some variation of this method and achieved results that were systematically limited at the 5 \% level \cite{Gri68,Str78,Byr02}. The method used by aCORN, first proposed by Yerozolimsky and Mostovoy \cite{Bal94,Yer04}, relies on a novel time-of-flight (TOF) asymmetry that does not require precise proton spectroscopy. The aCORN method is illustrated in figure \ref{F:cylfig}. Assume a pointlike cold neutron source on the axis between a set of opposing electron and proton detectors with a uniform axial magnetic field applied throughout. Electron and proton collimators, shown schematically as cylindrical tubes, lie on the axis. When a cold neutron, which is effectively at rest, decays, a beta electron, antineutrino, and proton are emitted. Due to their helical motion in the magnetic field $B$, the collimators impose a maximum transverse momentum of $p_{\perp\mbox{(max)}} = eBr/2$, where $r$ is the collimator radius, for detected electrons and protons. An electrostatic mirror containing a uniform axial electric field, produced by a pair of grids at ground and +3 kV as shown, causes all neutron decay protons to be accelerated and directed toward the proton detector. Electrons in the energy range of interest must be emitted into the right hemisphere to be detected. The momentum acceptances for the electron and proton in this scheme are shown in figure \ref{F:cylfig} (middle). These are cylinders in momentum space and the proton acceptance extends to both sides of the origin.  Now consider the antineutrino momentum acceptance for coincidence-detection events where the electron momentum is $\vec{p_e}$ as shown. Conservation of momentum requires $\vec{p_{\nu}} = - (\vec{p_e} + \vec{p_p})$ so the antineutrino momentum acceptance is a cylinder equivalent to the proton acceptance cylinder but displaced from the origin by $-\vec{p_e}$. If we neglect the kinetic energy of the proton (751 eV maximum) the electron and antineutrino must share the total decay energy $Q$ = 1293 keV and conservation of energy requires $|\vec{p_{\nu}}| = Q - \sqrt{p_e^2 + m_e^2}$. So for the given $\vec{p_e}$ the antineutrino momentum must lie on the intersection of the cylinder and sphere shown in figure \ref{F:cylfig} (bottom), which is indicated by the gray regions marked I and II. Region I (II) antineutrinos are correlated (anticorrelated) with $\vec{p_e}$ and have equal solid angles from the origin. If the $a$-coefficient is zero, the number of coincidence events associated with regions I and II will be equal. If not there will be an asymmetry. The same is true when we sum over all values of $\vec{p_e}$ for detectable electrons. In reality the neutron source is not a point but a cylindrical beam passing through the electrostatic mirror perpendicular to $\vec{E}$ and $\vec{B}$, so most decay vertices are off axis. For off-axis decays the proton and electron momentum acceptances are elliptical cylinders and the geometric construction is somewhat more complicated, but the result is essentially the same and solid angles of regions I and II remain equal.
\par
\begin{figure}
\centering
\includegraphics[width = 6.5in]{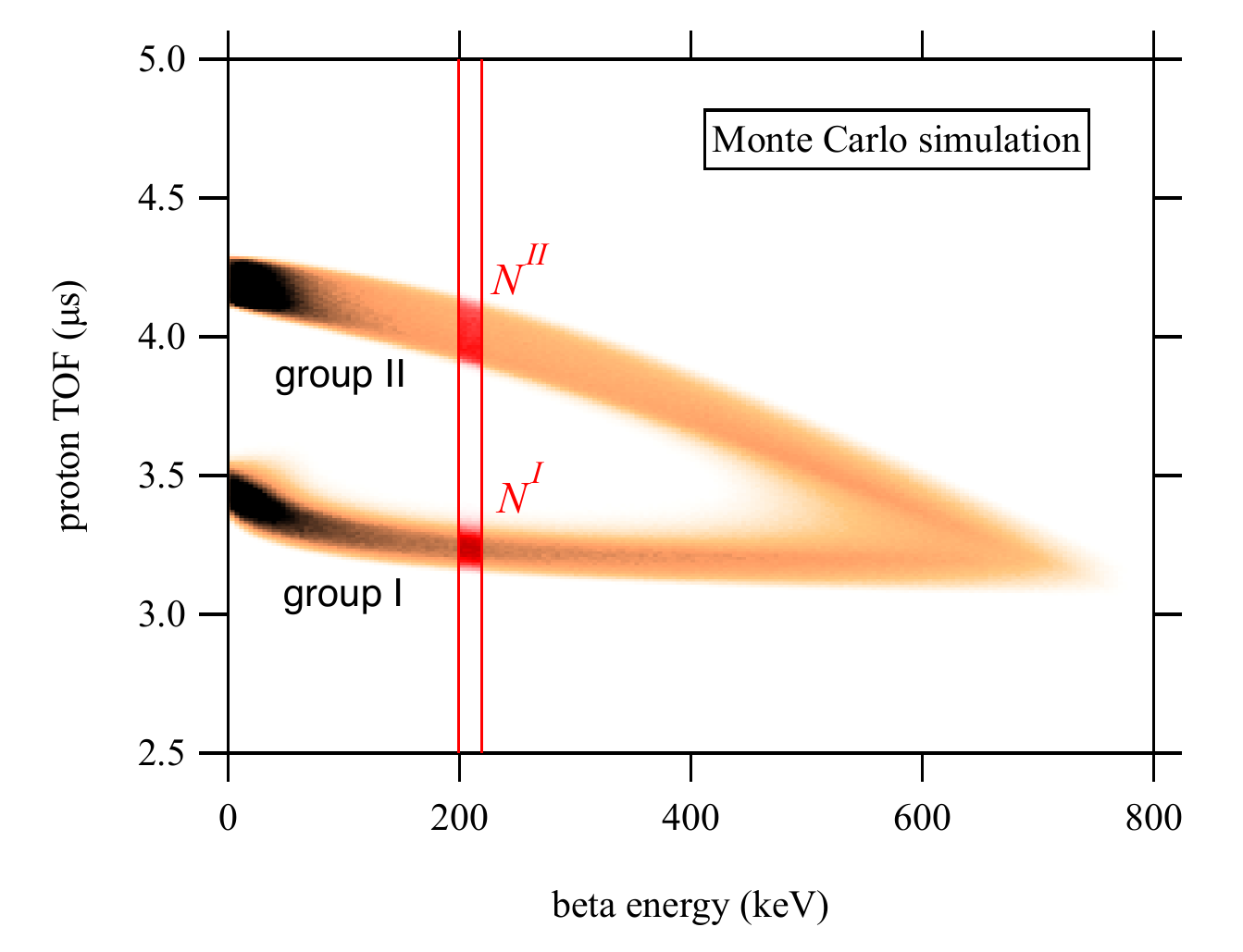}
\vspace{-0.2in}
\caption{\label{F:MCwishbone} A Monte Carlo simulation of aCORN data, proton TOF {\em vs.} beta energy for coincidence events. The fast proton branch (group I) is associated with neutron decays where the antineutrino momentum was in region I in figure \ref{F:cylfig}. The slow proton branch (group II) is associated with decays where the antineutrino momentum was in region II. The sums $N^I$ and $N^{II}$ are used to compute the wishbone asymmetry for each beta energy slice.}
\end{figure}

In the experiment we measure the beta electron energy and proton TOF, the time between electron and proton detection, for neutron decay events where both were detected. The data form the characteristic wishbone shape shown in figure \ref{F:MCwishbone}. Region I antineutrinos are correlated with the electron momentum direction, so the associated protons have larger momentum and axial velocity and the events lie on the lower wishbone branch (group I). Region II antineutrinos are anticorrelated with electron momentum, so the protons have smaller momentum and axial velocity and the events lie on the upper wishbone branch (group II). The gap between the wishbone branches corresponds to the kinematically forbidden gap between regions I and II in figure \ref{F:cylfig} (bottom). At beta energy above about 400 keV the regions overlap and the wishbone branches merge. A vertical slice at beta energy $E$, depicted in figure \ref{F:MCwishbone}, contains $N^I$ events in the lower branch and $N^{II}$ events in the upper branch. Using equation \ref{E:JTWeqn} we have
\begin{equation}
N^{I(II)}(E) = F(E) \int\int \left( 1 + a v \cos\theta_{e\nu} \right) d\Omega_e\, d\Omega^{I(II)}_{\nu}
\end{equation}
where $F(E)$ is the beta energy spectrum, $v$ is the beta velocity (in units of $c$), $\theta_{e\nu}$ is the angle between the electron and antineutrino momenta, and $d\Omega_e$, $d\Omega^{I(II)}_{\nu}$ are elements of solid angle of the electron and antineutrino (group I, II) momenta. The integrals are taken over the momentum acceptances shown in figure \ref{F:cylfig}. Since by construction the total solid angle products are equal for the two groups: 
$\Omega_e\,\Omega^I_{\nu}$ = $\Omega_e\,\Omega^{II}_{\nu}$, we find that the $a$-coefficient is related to the wishbone asymmetry $X(E)$ by
\begin{equation}
\label{E:asym}
X(E) = \frac{N^I(E) - N^{II}(E)}{N^I(E) + N^{II}(E)} = \frac{ \frac{1}{2} a v \left( \phi^I(E) - \phi^{II}(E) \right)}{1 + \frac{1}{2} a v \left( \phi^I(E) + \phi^{II}(E) \right)}
\end{equation}
The functions $\phi^I(E)$ and  $\phi^{II}(E)$ are defined as
\begin{equation}
\label{E:phis}
\phi^I(E) = \frac{ \int d\Omega_e \int_I d\Omega_{\nu} \cos\theta_{e\nu} }{\Omega_e \Omega_{\nu}^I} \quad\mbox{and}\quad
\phi^{II}(E) = \frac{ \int d\Omega_e \int_{II} d\Omega_{\nu} \cos\theta_{e\nu} }{\Omega_e \Omega_{\nu}^{II}},
\end{equation}
where again the integrals are taken over the momentum acceptances. Equations \ref{E:phis} can be understood as the average value of $\cos\theta_{e\nu}$ for detection regions I and II.  These are geometrical functions that depend only on the transverse momentum acceptances of the proton and electron so they can be calculated precisely from the known axial magnetic field and collimator geometries. 
\par
The second term in the denominator of equation \ref{E:asym} has a numerical value less than 0.005 in the energy range of interest (100 keV--380 keV), so we can treat it as a small correction and write
\begin{equation}
\label{E:aEffective}
X(E) = a f_a(E)\left[ 1 + \delta_1(E) \right] + \delta_2(E)
\end{equation}
with
\begin{equation}
\label{E:faE}
f_a(E) = \frac{1}{2} v \left( \phi^I(E) - \phi^{II}(E) \right)
\end{equation}
and 
\begin{equation}
\delta_1(E) = -\frac{1}{2} a v \left( \phi^I(E) + \phi^{II}(E) \right).
\end{equation}
The other small correction $\delta_2(E)$ in equation \ref{E:aEffective} comes from our neglect of the proton's kinetic energy in the momentum space discussion of figure \ref{F:cylfig}. If we account for this energy, the antineutrino momentum sphere is slightly oblong and the solid angles of groups I and II differ by approximately 0.1 \%. This causes a small (about 1 \% relative) intrinsic wishbone asymmetry that is independent of the $a$-coefficient; it is straightforward to compute by Monte Carlo to the needed precision.
\par
\begin{figure}
\centering
\includegraphics[width = 6.5in]{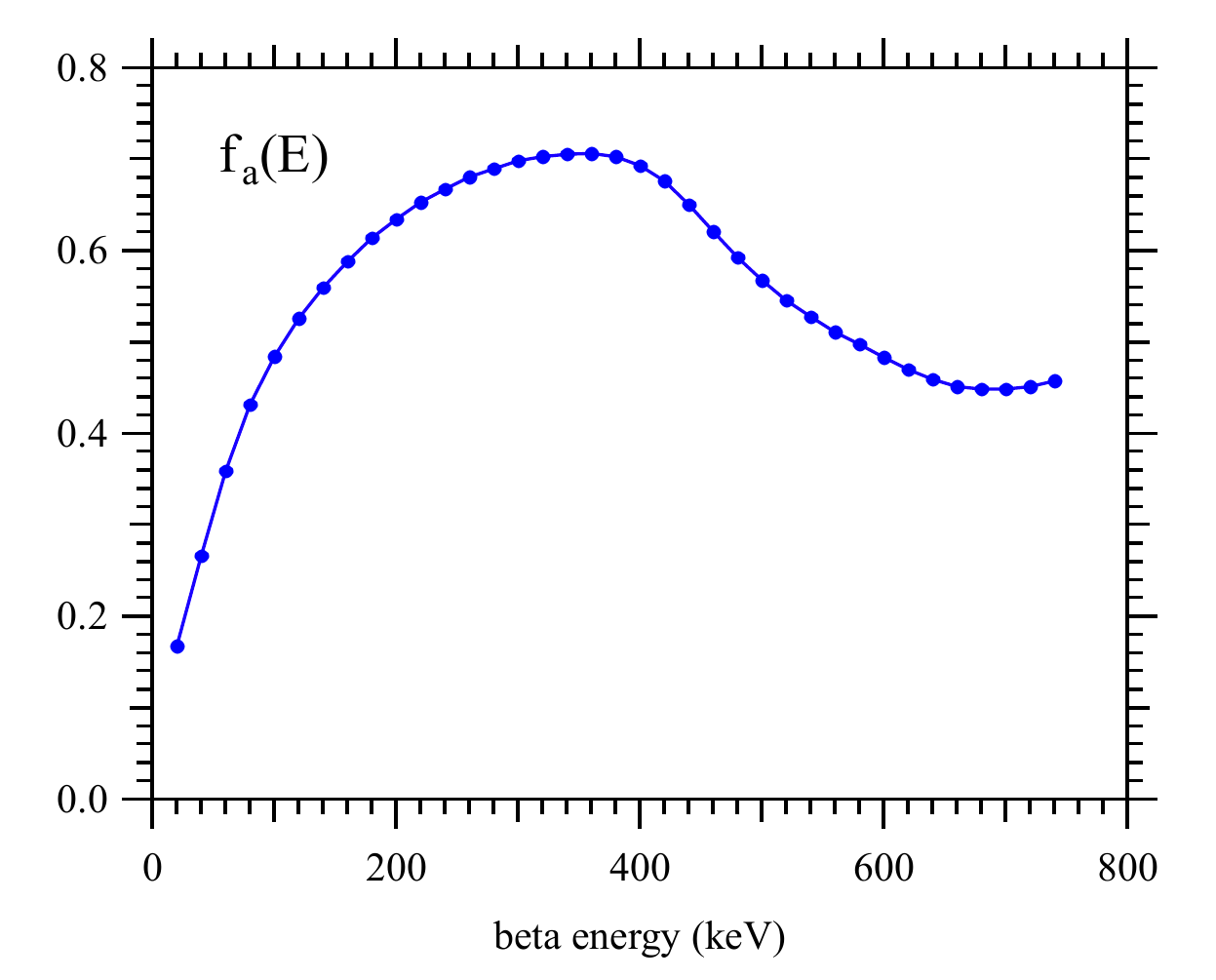}
\caption{\label{F:faE} The dimensionless geometric function $f_a(E)$, computed numerically from the aCORN geometry and a 36.4 mT uniform magnetic field (see equations \ref{E:aEffective}, \ref{E:faE}).}
\end{figure}

Omitting the small corrections we see that $X(E) = a f_a(E)$; the experimental wishbone asymmetry is proportional to the $a$-coefficient and the dimensionless geometric function $f_a(E)$. In analyzing the data we take the approach of assuming a perfectly uniform axial magnetic field and exact collimator configuration, and use the precisely computed $f_a(E)$ shown in figure \ref{F:faE}. We then treat nonuniformities and uncertainties in the measured magnetic field magnitude and shape and the collimator geometry as systematic effects applied to the result.
\par
aCORN runs on a nominally unpolarized neutron beam. If the beam were slightly polarized, there would be an additional contribution to the wishbone asymmetry from the antineutrino asymmetry correlation $B$ term in equation \ref{E:JTWeqn}, giving
\begin{equation}
\label{E:XEpol}
X(E) = a f_a(E) + PB f_B(E)
\end{equation}
where $P$ and $B$ are the neutron polarization and $B$-coefficient, and $f_B(E)$ is a similarly calculated geometric function for $PB$. Because neutron polarization is more axially peaked than 
${\bf p}_{\bf e}$, $f_B(E)$ is on average 40 \% larger than $f_a(E)$. Also $|B/a| \approx 10$. So even with $P\ll 1$ this can be a significant effect. In the NG-6 aCORN data an observed difference in $X(E)$ with the magnetic field in the up and down directions was attributed to a neutron polarization $P \approx$ 0.6 \% \cite{Dar17}.

\section{The \lowercase{a}CORN Apparatus}
\label{S:apparatus}
\begin{figure}
\centering
\includegraphics[width = 5in]{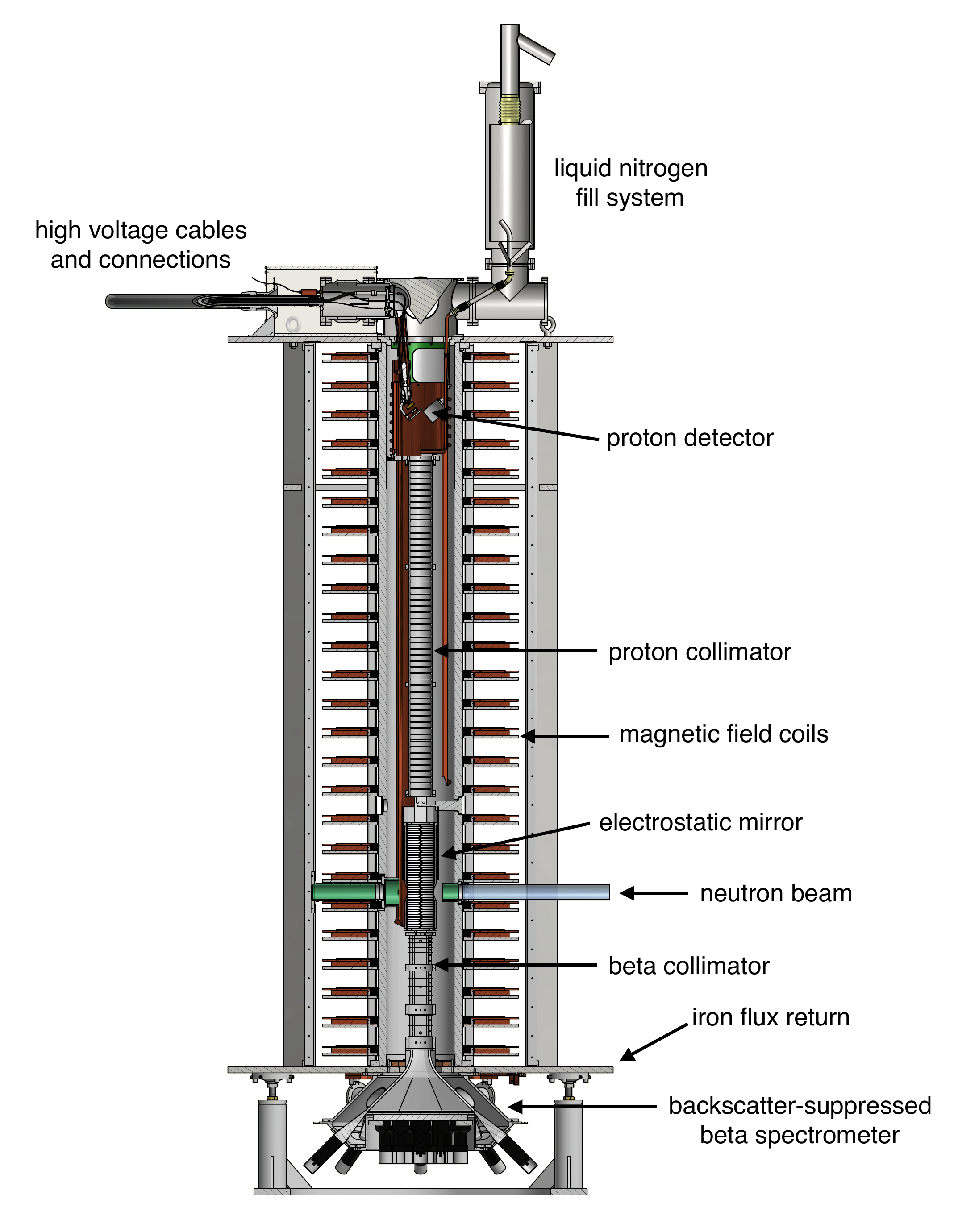}
\caption{\label{F:tower} A cross section view of the aCORN tower showing the arrangement of major components. The neutron beam passes through from right to left.}
\end{figure}

We describe here briefly the main components of the aCORN apparatus. More details can be found in previous publications \cite{Wie09,Col17,Has17}. Figure \ref{F:tower} shows a cross section view of the aCORN tower.
\par
The 36.3 mT axial main magnetic field is produced by a vertical array of 24 individual flat coils supplied in series. Each coil contains 121 turns of 2 cm $\times$ 0.1 cm copper tape, has an overall diameter of 78.8 cm, and rests on a water-cooled copper plate. Coil assemblies are separated by 8 cm vertical gaps, set by the size of the neutron beam. The full magnet assembly is surrounded by an iron flux return yoke composed of top and bottom circular endplates and four vertical columns. A set of 76 computer controlled trim coils are used to improve the shape of the magnetic field. Each main coil has an attached axial trim coil. Two pairs of large transverse coils cancel the overall environmental transverse field. Twenty-four pairs of small transverse trim coils are used to eliminate localized transverse fields and gradients. A robotic magnetic field mapper, attached to precision bearings on the upper and lower iron endplates, is used to map the magnetic field inside the vacuum chamber, both on and off axis. Using the results of these maps, an algorithm computes the trim coil currents needed to meet the magnetic field specifications. The proton and electron collimators, and the electrostatic mirror, are then optically aligned to the axis of the experiment defined by the bearings.
\par
The electron collimator is a series of seventeen 0.5-mm thick tungsten discs, each with a 5.5 cm diameter circular aperture. These are unevenly spaced to minimize the probability that an electron will scatter from an edge and reach the active area of the beta spectrometer, as determined by a PENELOPE simulation. The total length of the electron collimator is 48.0 cm. The proton collimator is a monolithic aluminum tube, 140.0 cm long, containing  a series of 49 evenly spaced 8.0 cm diameter knife edge apertures cut by a precision lathe on the inner surface. The electron and proton collimators are individually aligned and attached to a rigid aluminum insert structure which is then aligned as a single unit to the experimental axis.
\par
The electrostatic mirror must provide a nearly uniform axial electric field in the cylindrical neutron decay region. This requires differing uniform potentials at the ends and a linearly varying potential on the wall. The neutron beam must also penetrate the mirror wall, which presented a technical challenge. Our solution was to make the wall from a thin (0.25 mm) polytetrafluoroethylene (PTFE) sheet. The inner surface of the sheer was electroplated with a 4.5 $\mu$m layer of copper divided into 63 parallel thin bands by photolithography, produced by Polyflon \cite{Polyf, DISCLM}. These 63 bands were held at potentials established by a chain of equal precision resistors to approximate the linear boundary condition. The neutron beam was allowed to pass through the wall on both sides, each side scattering about 1 \% of the beam by the PTFE and scattering/absorbing about 0.1 \% of the beam by the copper. The end potentials were set by grids of 100 $\mu$m wires. The grid on the bottom (electron) side was at +3 kV and the grid on the top (proton) side was at ground.
\par
\begin{figure}
\centering
\includegraphics[width = 5in]{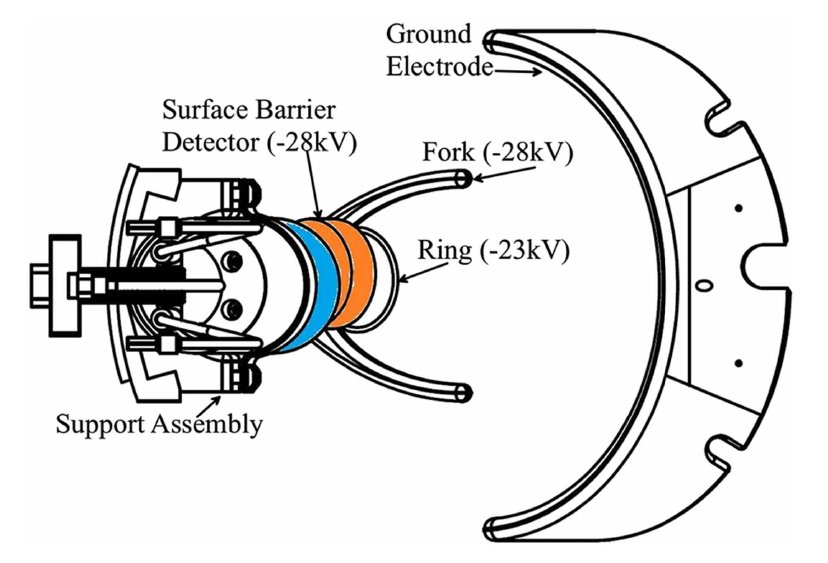}
\caption{\label{F:protAssy} An overhead view of the proton detector assembly showing the positions of the surface barrier detector and focusing electrodes.}
\end{figure}
\begin{figure}
\centering
\includegraphics[width = 5in]{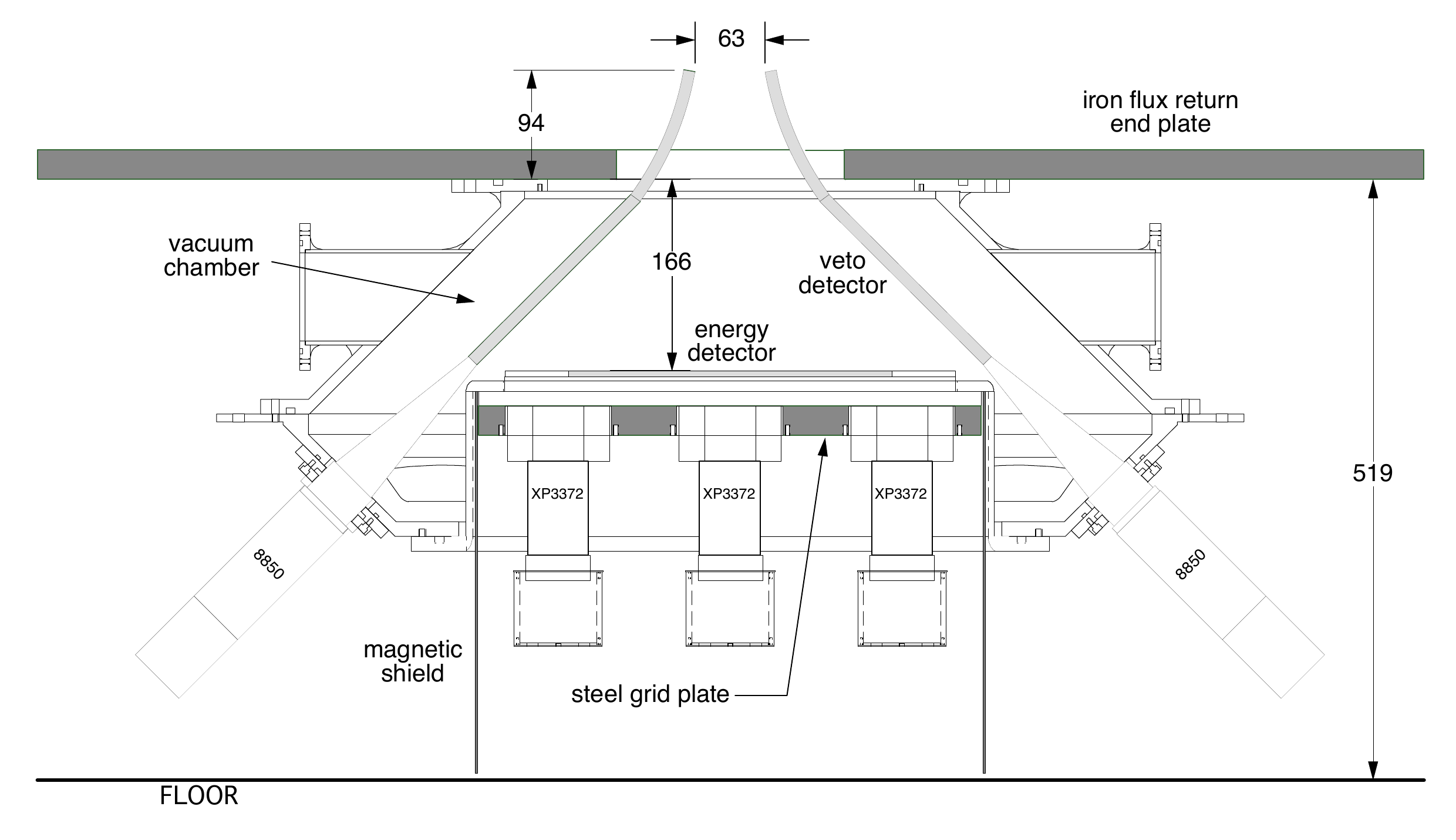}
\caption{\label{F:bsCutaway} An interior cross section view of the backscatter-suppressed beta spectrometer. Dimensions are in mm.}
\end{figure}

The proton detector was a 600 mm$^2$, 1000 $\mu$m thick surface barrier detector held at -28 kV to accelerate protons to a detectable energy. Figure \ref{F:protAssy} shows an overhead view, looking down from the top of the tower. Detector components were located off-axis to prevent neutron decay electrons emitted in the upward direction from backscattering on the proton detector and returning to the beta spectrometer, where they would be detected with the wrong energy and wrong sign of $\cos\theta_{e\nu}$. A focusing fork and ring act as a lens to focus all protons exiting the proton collimator onto the active area of the detector. The proton detector is cooled by a copper panel attached to a liquid nitrogen cooling system. 
\par
aCORN employed a novel backscatter-suppressed beta spectrometer, illustrated in cross-section in figure \ref{F:bsCutaway}. The beta energy detector was a 5 mm thick, 280 mm diameter circular slab of Bicron BC-408 plastic scintillator, viewed by 19 Photonis XP3372 8 stage 7.6 cm (3 inch) hexagonal photomultiplier tubes (PMT's). Surrounding the energy detector was an array of eight veto detectors, each composed of a 10 mm thick BC-408 plastic scintillator and adiabatic acrylic light guide viewed by a Burle 8850 12 stage 5.1 cm (2 inch) PMT. The spectrometer was mounted on the tower below the bottom flux return end plate. The axial magnetic field was high at the entrance to the spectrometer but dropped quickly below it to about 1 mT at the energy detector. All beta electrons with kinetic energy $>$100 keV that were accepted by the beta collimator passed through the opening at the top of the veto array and struck the active area of the energy detector, as verified by Monte Carlo simulation. Approximately 5 \% were expected to backscatter from the plastic scintillator without depositing their full energy. This may lead to a large systematic effect, discussed in section \ref{SSS:backscatter}. To mitigate this a backscatter veto array was used; the majority of backscattered electrons struck a veto paddle and were vetoed. The overall veto efficiency for backscattered electrons was measured to be ($92\pm5$) \%. A pair of linear motion vacuum feedthroughs located between the electrostatic mirror and proton collimator held conversion electron sources ($^{113}$Sn and $^{207}$Bi). During production runs, {\em in situ} calibration measurements were made at approximately 48 hour intervals to monitor slow gain drifts in the beta spectrometer and enable correction in the data analysis. Details of the design, construction, and characterization of the aCORN beta spectrometer can be found in reference \cite{Has17}.
\par
The main vacuum chamber of aCORN was a vertical aluminum tube 3 m tall and 28 cm inner diameter. It was joined at the top and bottom to the iron endplates by o-ring seals. A 250 l/s turbomolecular pump was mounted on the beta spectrometer chamber and a 370 l/s helium cryopump was attached to the beam dump. A set of three liquid nitrogen cooled copper cryopanels extended from the top of the main chamber to the bottom of the proton collimator to provide high conductance pumping of water and volatiles released by the plastic scintillator in the beta spectrometer. During normal operation the pressure at the top of the electrostatic mirror was about $8 \times 10^{-5}$ Pa ($6 \times 10^{-7}$ torr).

\section{Modifications for the NG-C run}
\label{S:NGC}
A previous publication \cite{Col17} describes the aCORN apparatus as it was used for the first measurement on the NG-6 beamline at the NIST Center for Neutron Research (NCNR)  \cite{NCNR} in 2013--2014. The experiment was moved, with some modifications, to the new high-flux beamline NG-C in 2015 for a second run. This section describes those modifications.
\subsection{Neutron beam and collimation}
In 2013 a second guide hall was commissioned at the NCNR with four new supermirror guides. The end position on new guide NG-C was designated for fundamental neutron physics experiments and aCORN was the first experiment to run there. NG-C is a ballistic curved supermirror guide 11 cm $\times$ 11 cm at the exit with a measured capture flux of $8.1\times 10^9$ cm$^{-2}$s$^{-1}$. Details of the design of the NG-C guide and other guides in the new guide hall can be found in \cite{Co09}. Because NG-C is curved, a bismuth filter is not needed to remove fast neutrons and gammas, which improves neutron transmission to the experiment. A 180-cm long secondary focusing supermirror guide was installed to reduce the beam cross section to a 6 cm $\times$ 6 cm square. This was followed by a neutron collimator, 120 cm long containing four $^6$LiF apertures. Its interior was lined with $^6$Li glass to absorb scattered neutrons. The collimator reduced the beam divergence and delivered a 3.1-cm diameter circular beam to the experiment. The capture flux in the neutron decay region of aCORN was measured to be $6.7\times 10^9$ cm$^{-2}$s$^{-1}$, about a factor of ten higher than the equivalent measurement with the experiment installed on NG-6, but with a beam area that was a factor of two smaller, resulting in an overall factor of five increase in the wishbone event rate from neutron decay. 
\par
At the end of NG-C is a 2.4-m deep pit available to experiments that need part of the apparatus below floor level. For aCORN we constructed a false floor inside the pit at 40 cm below the main floor level for better access to the beta spectrometer and the field mapping apparatus when installed.
\subsection{Electrostatic mirror}
\begin{figure}
\centering
\includegraphics[width = 5in]{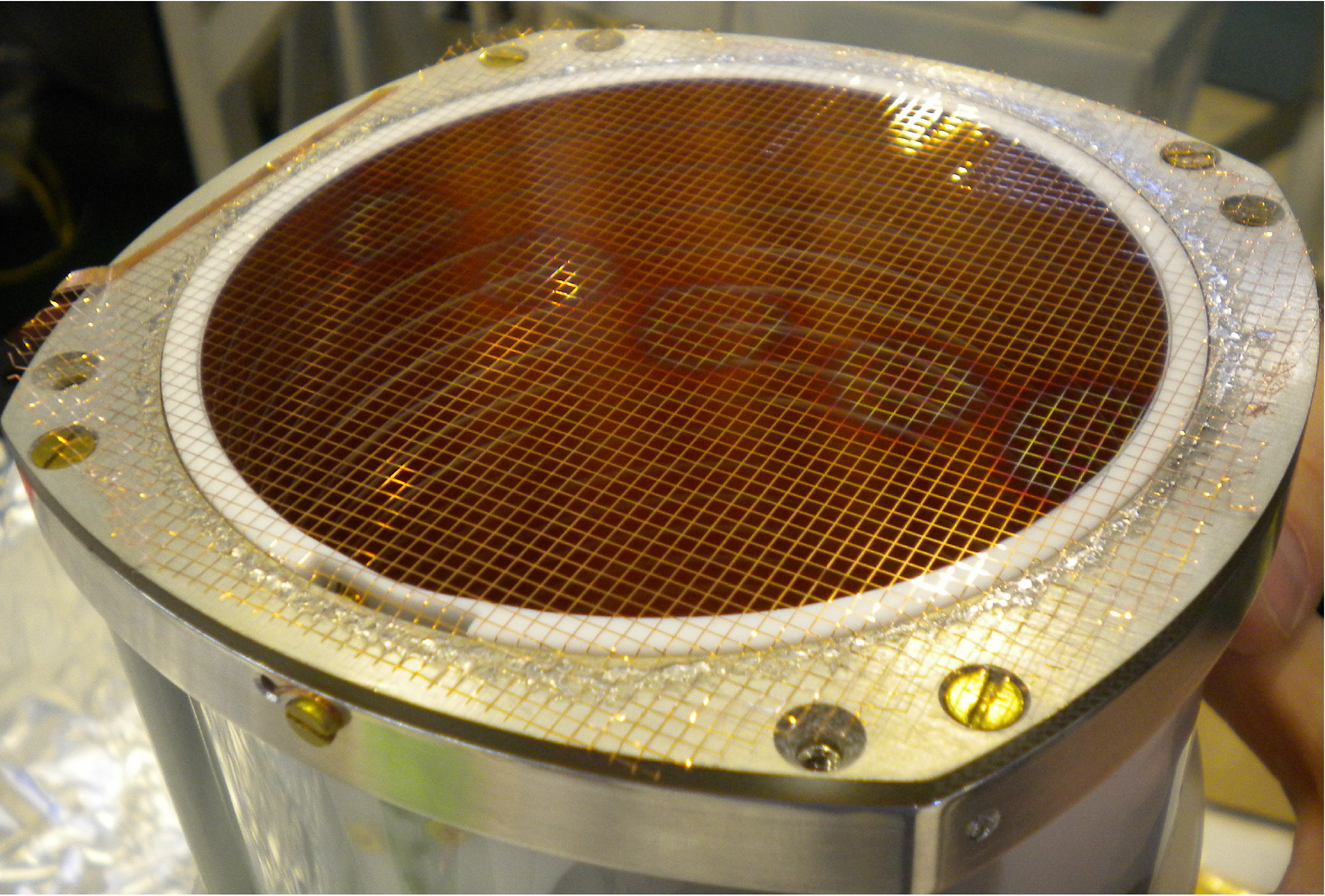}
\caption{\label{F:newGrid} The redesigned upper end of the electrostatic mirror used in the NG-C run, showing the new square mesh grid and larger open diameter.}
\end{figure}

The departure from a perfectly axial electric field in the vicinity of the upper grounded end grid, where the protons pass through, resulted in the largest systematic correction (\mbox{5.2 \%}) and 
uncertainty (\mbox{1.1 \%}) in the result from the NG-6 run \cite{Dar17}. Guided by a 3D COMSOL \cite{COM} model along with a Monte Carlo proton transport simulation, we made some improvements to the upper grid geometry to reduce this effect. We replaced the linear wire upper grid with an electroformed square mesh copper grid containing 100 $\mu$m threads spaced by 2 mm, purchased from Precision Eforming \cite{PE}. We also redesigned the upper aluminum support ring to locate it entirely outside the thick PTFE tube, thereby increasing the open inner diameter at the top to 10.9 cm. The new upper grid can be seen in the photo in figure \ref{F:newGrid}. These adjustments reduced the size of the electrostatic mirror correction by more than a factor of three (see the discussion in section \ref{SSS:ESmirror}). The lower +3 kV grid was left unchanged; protons do not pass close to the lower grid and the electrostatic effect on electrons passing through it is negligible.
\subsection{Data Acquisition}
\label{SS:daqNGC}
Electronic pulses from the 19 beta energy channels, 8 backscatter veto channels, and the proton detector were sent to two PIXIE-16 modules \cite{XIA} which are 12 bit, 100 MSPS multiplexing analog to digital converters. For the NG-C run we made two changes to the PIXIE-16 firmware: i) certain calculations that were not needed, such as constant fraction discrimination ratios, were removed in order to increase throughput; and ii) the energy calculation for all channels was switched from a trapezoidal filter to the charge to digital conversion (QDC) mode. In the QDC mode three timings are specified: 1) time before the event trigger to begin saving data (0.6 $\mu$s for electrons, 1 $\mu$s for protons), 2) the pre-pulse time window (0.5 $\mu$s for electrons, 0.5 $\mu$s for protons), and 3) the pulse time window (0.3 $\mu$s for electrons, 2 $\mu$s for protons). The energy is then calculated as the average number of counts per channel in the pulse window minus the average number of counts per channel in the pre-pulse window. We found that this switch did not noticeably affect energy resolution or linearity, but it significantly lowered the effective energy threshold which was useful for all channels but was particularly helpful for the proton channel. In the NG-6 data analysis \cite{Dar17} there was a \mbox{3 \%} systematic correction to the $a$-coefficient due to loss of protons below threshold. Such a correction was not needed in the NG-C data analysis.

\section{The NG-C run}
aCORN ran on the NG-C end position at the NCNR from August 2015 to September 2016 and collected a total of 3758 beam hours of neutron decay data. The raw coincidence event rate was 171 s$^{-1}$, The neutron decay wishbone event rate, after background subtraction, was 0.9 s$^{-1}$, about a factor of three higher than in the previous run on NG-6. 
\par
\begin{figure}
\includegraphics[width = 5in]{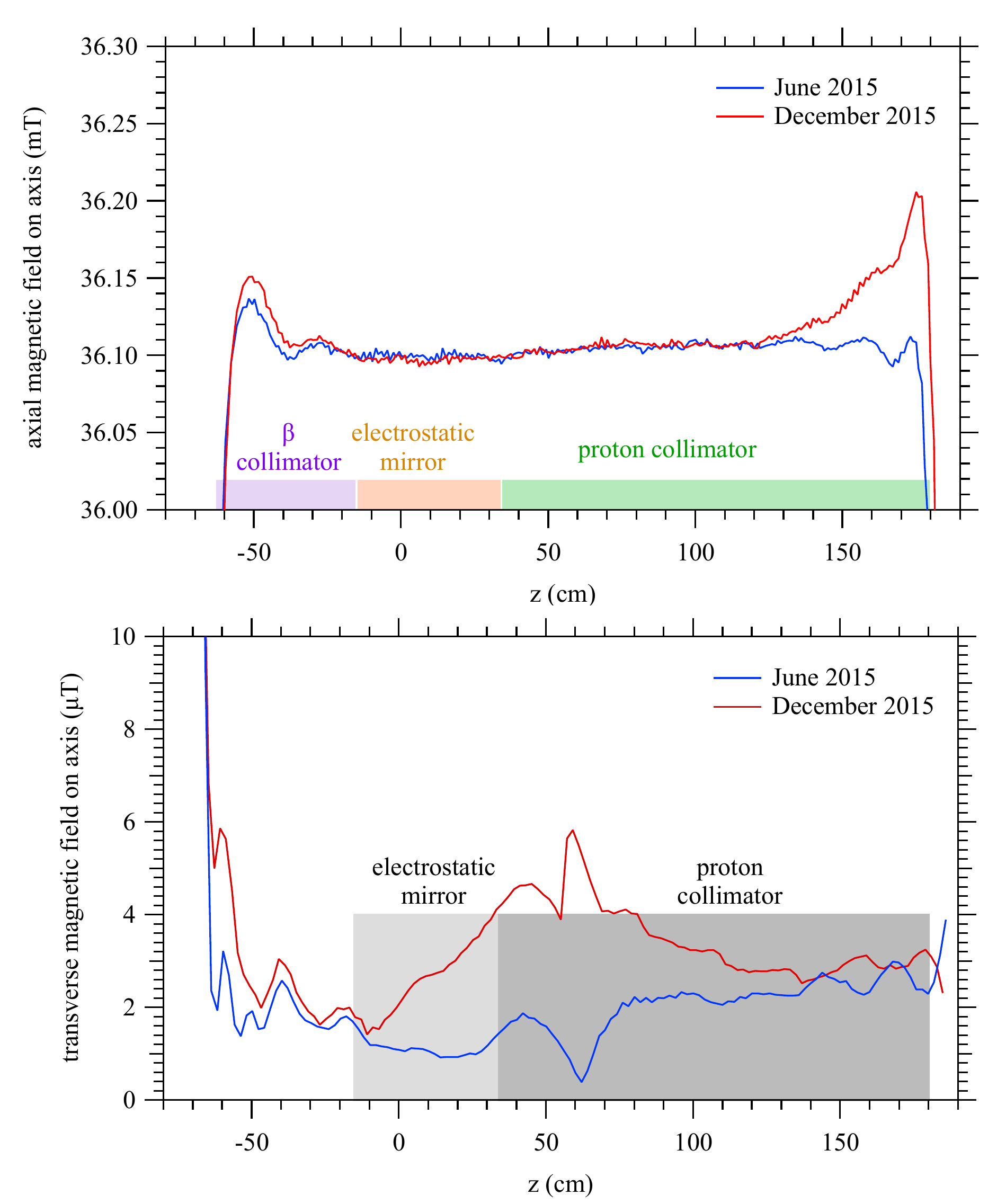}
\caption{\label{F:bMapsAxis} The axial (top) and transverse (bottom) magnetic fields measured by the robotic field mapper on axis. The June 2015 maps were made after reversing and
trimming the field.  The December 2015 maps were made just prior to the next field reversal. The difference shows typical drift over six months with unchanged trim coil settings. Gray shaded regions
in the bottom plot indicate the $<4$ $\mu$T target for the transverse field in the electrostatic mirror and proton collimator.}
\end{figure}

We collected data in both axial magnetic field directions in order to monitor and correct for a possible effect due to residual polarization of the neutron beam. The first data run was magnetic field up (B$_{\rm up}$), for 1097 beam hours. The second run was magnetic field down (B$_{\rm down}$), for 2178 hours. The magnetic field was returned to B$_{\rm up}$ for the final 482 hours. The following protocol was followed whenever the magnetic field was reversed: 1) the detectors and collimation insert were removed and the field mapper installed; 2) the existing axial and transverse magnetic fields were mapped and compared to the previous maps; 3) the leads to the main magnet supply were reversed and all trim coils were de-energized; 4) the magnetic field was mapped and trimmed to specification in the new direction; 5) the field mapper was removed and the detectors and collimation insert reinstalled and aligned. The entire process of reversing the magnetic field took about two weeks, completed mostly during NCNR refueling shutdown periods. Figure \ref{F:bMapsAxis} shows results of on-axis axial and transverse field maps made in June 2015, prior to the first production run, and December 2015, just before the first field reversal and with the trim settings unchanged. Drifts in the field shape over the six month span are evident in the plots. We attribute the increase in axial field near the top and bottom of the tower to relaxation of the flux return endplates. Our target uncertainty for the axial field is $\pm0.2$ mT so this axial drift is not a problem. aCORN is very sensitive to transverse magnetic fields in the proton transport region, {\em i.e.} the electrostatic mirror and proton collimator, as they can cause a false wishbone asymmetry. As can be seen in figure \ref{F:bMapsAxis} (bottom) the newly trimmed field in June met our target of $<4$ $\mu$T, but in December the transverse field in a region near the bottom of the proton collimator exceeded the target. However the associated systematic effect was small (see section \ref{SSS:magfield}). 
\par
\begin{figure}
\centering
\includegraphics[width = 5in]{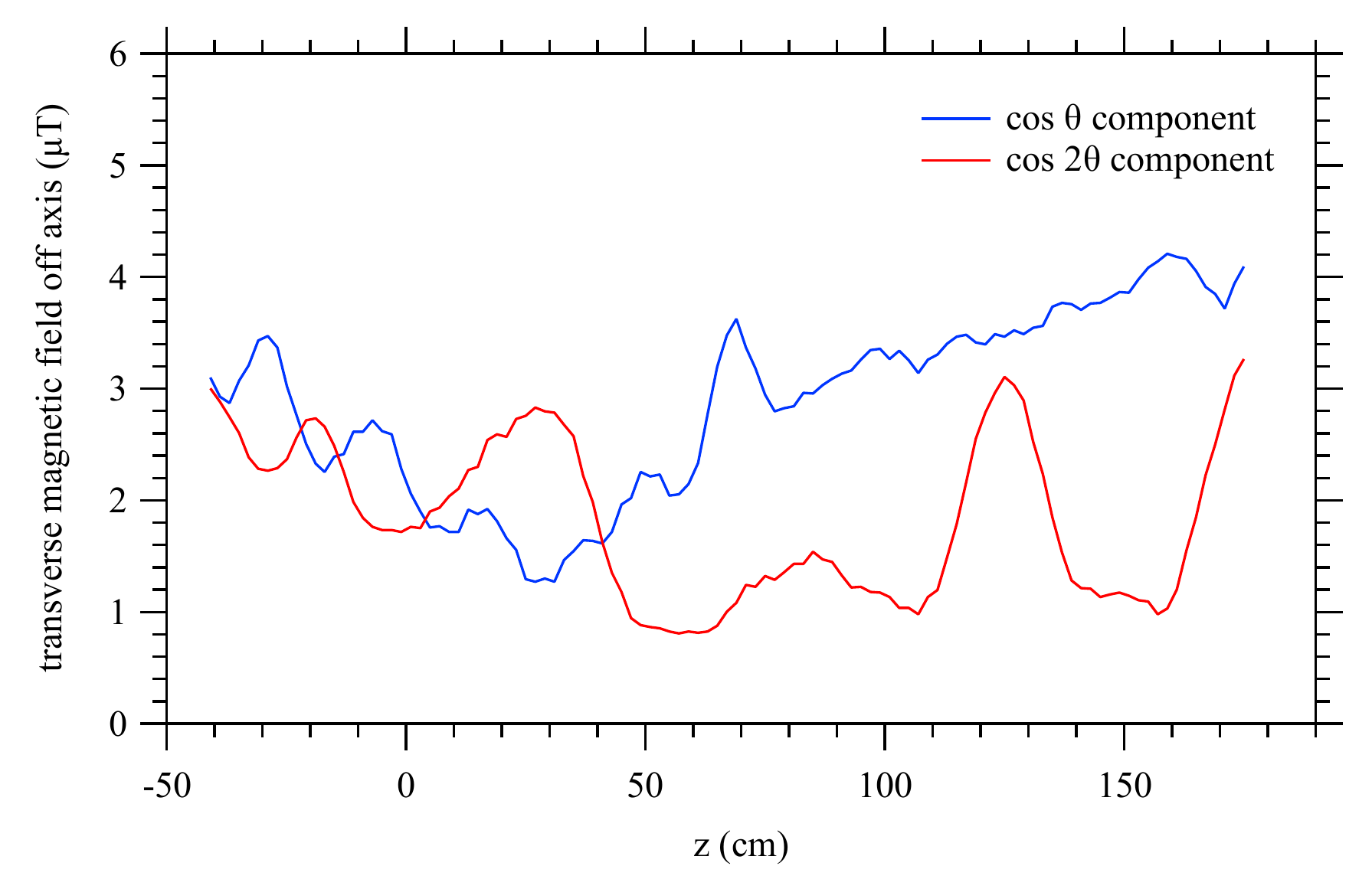}
\caption{\label{F:bMapOffAxis} The transverse field map measured 5.1 cm off axis using the robotic field mapper. At each $z$ position the field is measured at 30$^{\circ}$ intervals as the
mapper rotates. The result is Fourier decomposed into a $\cos\theta$ component that gives the uniform transverse field and a  $\cos2\theta$ that corresponds to a transverse gradient.}
\end{figure}

Figure \ref{F:bMapOffAxis} shows a transverse field map taken 5.1 cm off-axis. At each $z$ position the field was measured in steps of 30$^{\circ}$ as the mapper carriage rotated. The data were fit to a Fourier series function: $B_{\rm trans}(\theta) = b_0 + b_1 \cos(\theta - \theta_1) + b_2 \cos2(\theta - \theta_2)$. The constant term $b_0$ is dominated by the small misalignment angle between the Hall probe and the field axis and is not interesting. The $\cos\theta$ coefficient $b_1$ gives the uniform transverse field off axis. The $\cos2\theta$ coefficient $b_2$ results from a transverse gradient.  The parameters $\theta_1$ and $\theta_2$ are constant phase offsets.
\par
Figure \ref{F:insertAlign} shows results of alignment checks of the collimation insert made at various times during the run. Measurements were made using an optical system consisting of a theodolite, a pentaprism that rotates the line of sight by 90$^{\circ}$, and a series of precision reticules installed in the insert, all in a very well measured geometry. Usually independent measurements were made by two people as a double-check. The electrostatic mirror alignment was consistently within our target of 1 mrad. The proton collimator had a much stricter target of 0.1 mrad which was generally met or slightly exceeded.

\section{Data Reduction and Analysis}

For each aCORN event the PIXIE system recorded the energy and time of 31 signals: 19 beta energy PMTs, 8 beta veto PMTs, 2 copies of the proton preamp output, and 2 copies of a level-discriminated 
proton pulse. An event was defined in firmware as any two of the above signals above threshold within a 100 ns time window. Two copies of the proton detector signal were used so that a single proton would produce an event. Most noise and dark current from individual PMTs did not produce events. These raw data were written to disk along with header information containing run parameters at a rate of about 5 TB per day. An online data distiller preprocessed raw data and removed much of the background. The distiller included all events that were within a time window 10 $\mu$s before to 1 $\mu$s after each proton event. Events outside this window could not be a neutron decay coincidence and were discarded. Data bottlenecks within the PIXIE could cause events to enter the data stream out of time order, but each event contained an accurate time stamp used by the distiller to correct the time order. The distiller produced distilled data files at a rate of about 8 GB per day (a factor of $>$600 reduction) that became the archival data. Raw data were not saved, except for a small sample kept each day for diagnostic purposes. 
\par
A data reducer was then used to convert the distilled data into reduced data files, individual text files each containing 160 s of coincidence event data, for analysis. The reducer combined individual beta PMT events into complete beta energy and time, or discarded them as noise, assigned a veto state to each, and calculated the beta-proton time of flight (TOF) for each proton event within the 11 $\mu$s time window. The reduced data were organized into series of up to 1000 files, about two days of data, collected under essentially the same experimental conditions. Each series had an associated beta energy calibration obtained from {\em in situ} calibration source measurements completed every two days.
\par
\begin{figure}
\centering
\includegraphics[width = 6in]{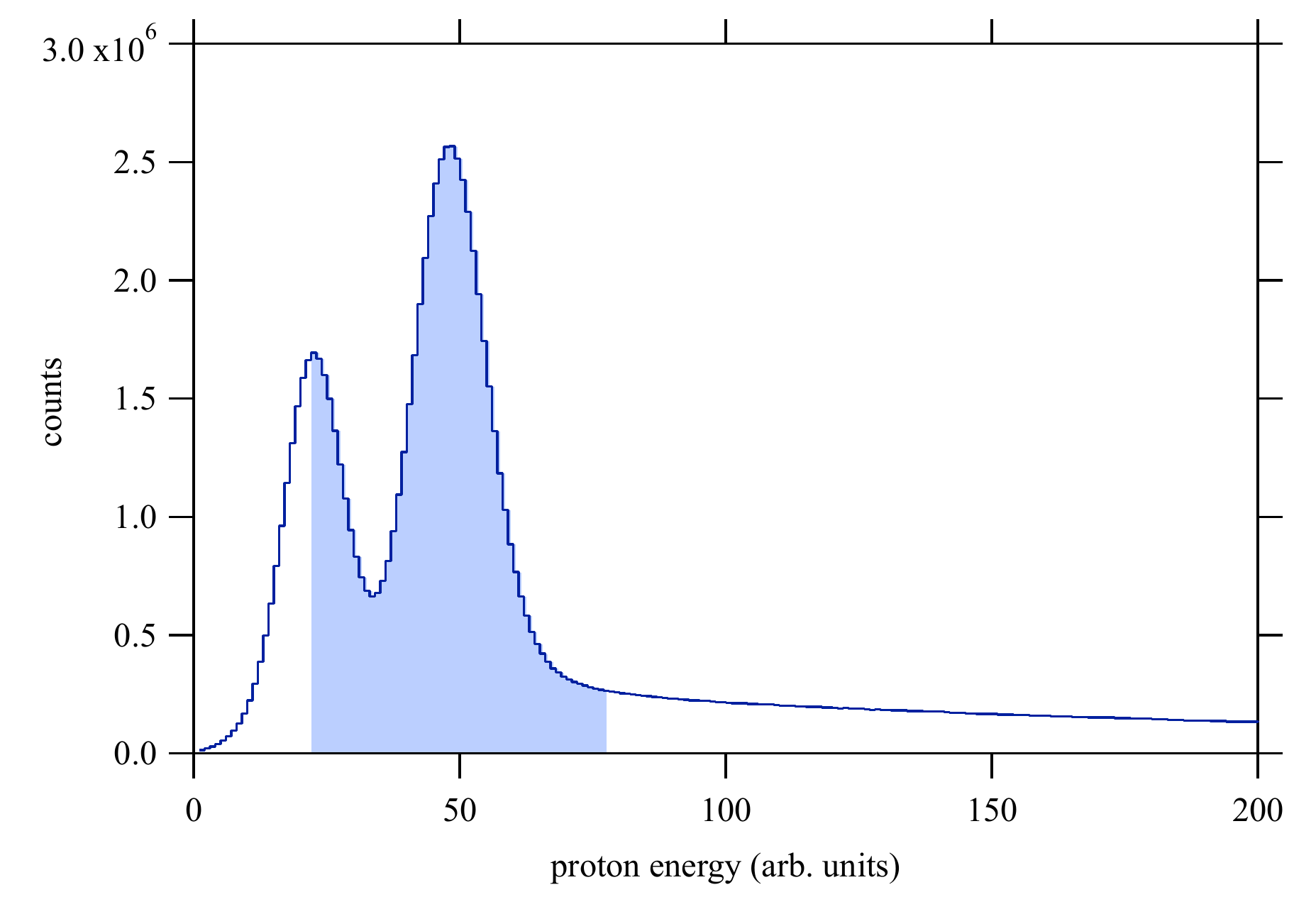}
\caption{\label{F:pCut} A typical proton energy singles spectrum. The peak on the right is protons. The noise/background forms a peak on the left due to the soft energy
threshold of the PIXIE. The shaded region is the applied proton energy window.}
\end{figure}

Data were divided into groups, each containing several equivalent series totaling approximately 100 beam hours, for analysis. Data were then sorted into a raw wishbone plot, a plot of proton TOF {\em vs.} beta energy, applying the calibration data for each series separately, with a proton energy cut applied as shown in figure \ref{F:pCut}. A typical raw wishbone plot is shown in figure \ref{F:wishRaw}. Neutron decays are contained in the ``wishbone'' structure of delayed coincidence events.

\subsection{Data Blinding Strategy}
The nature of aCORN does not allow an easy way to add an arbitrary blinding constant to the wishbone data. But the possibility of residual neutron polarization offers a useful data blinding strategy. An unknown neutron polarization would add an offset to the wishbone asymmetry, as shown in equation \ref{E:XEpol}, that is undetectable in the analysis of data from a single magnetic field direction. In the NG-6 run 
a presumed neutron polarization of only 0.6 \% produced an 8.4 \% shift in the value of the $a$-coefficient for each field direction \cite{Dar17}. Our blinding strategy was as follows:
\begin{enumerate}
\item A small subset of the aCORN collaboration, the polarimetry group, measured the aCORN neutron beam polarization {\em in situ} using polarized $^3$He NMR in an auxiliary experiment and analyzed the result, which was not revealed to other collaboration members.
\item The magnetic field up (B$_{\rm up}$) data only were fully analyzed, including all systematic corrections and uncertainties, and a result for the $a$-coefficient was obtained and locked. The polarimetry group did not participate in this analysis.
\item The magnetic field down (B$_{\rm down}$) data were analyzed using the same procedures and corrections, without adjustment.
\item The polarimetry ``box'' was opened and the result compared to the $a$-coefficients from the B$_{\rm up}$ and B$_{\rm down}$ analyses.
\end{enumerate}

\subsection{Background Subtraction and Dead Time Correction}
The PIXIE system is complicated and exhibits dead time effects at several time scales. First there is a dead time for each  channel that depends on the time structure of its signal pulses. This was 300 ns for beta PMT channels and 3000 ns for proton detector channels. Because the analog to digital conversion is multiplexed, an additional deadtime of several $\mu$s can occur for a group of channels on a single module when data rates are high. Finally, when the module memory is full, the entire module of 16 channels is dead for about 3 $\mu$s while data is transferred to the host computer. During the NG-6 run the data rate was sufficiently low that the longer dead times were not apparent in the data, but during the NG-C run, with about five times higher data rate, such affects did appear. We found that a 4 $\mu$s dead time for all events, applied in the analysis, was sufficient to remove all nonphysical time correlation effects between channels in the data. 
\par
\begin{figure}
\centering
\includegraphics[width = 6in]{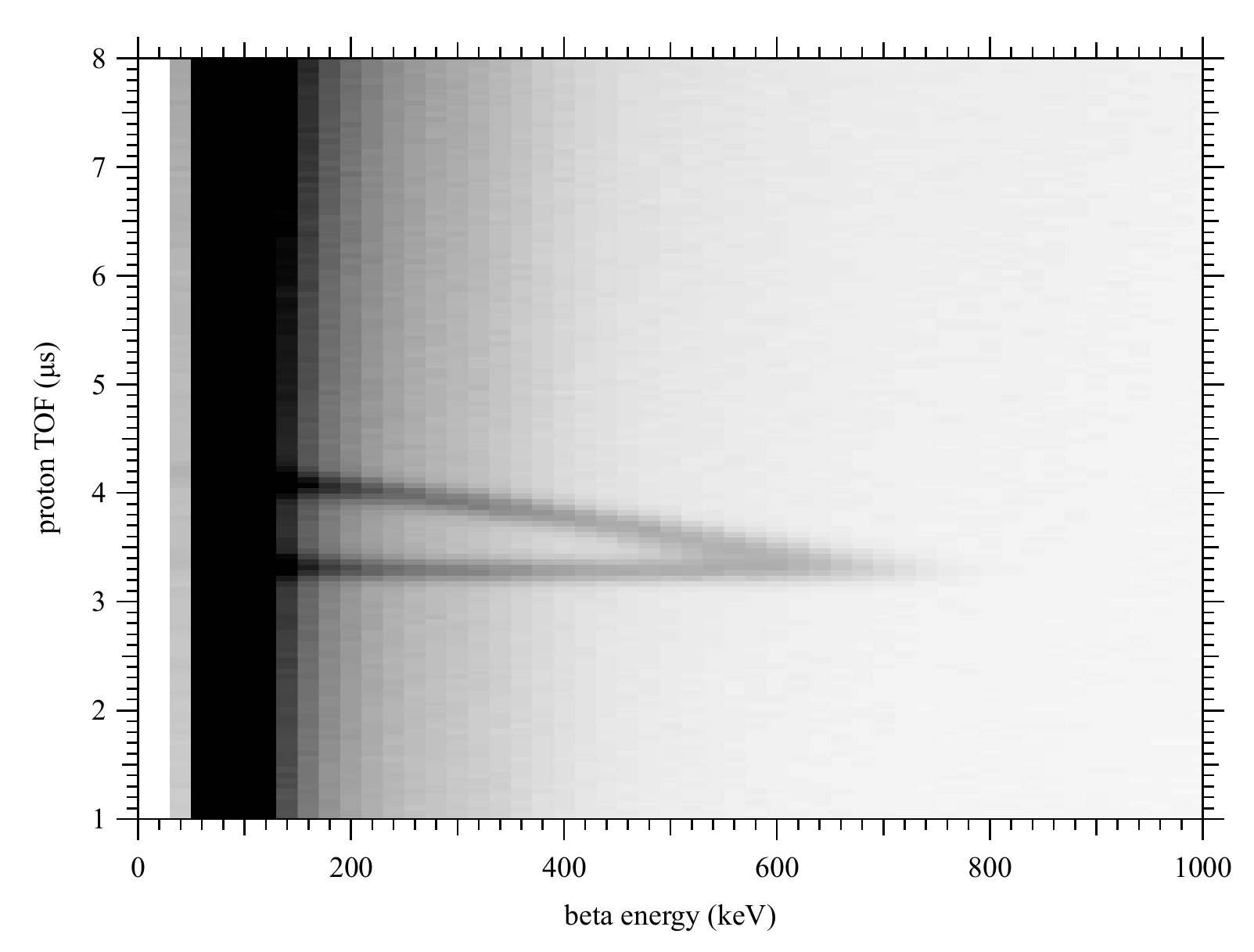}
\caption{\label{F:wishRaw} A typical raw wishbone obtained from approximately 100 hours of reduced data, using the proton energy cut shown in figure \ref{F:pCut}.}
\end{figure}

In the NG-6 data analysis described in \cite{Col17,Dar17} we were able to treat each electron event within the coincidence time window of a proton (10 $\mu$s before the proton to 1 $\mu$s after) as a separate coincidence event. The same proton could be associated with several different coincidence events, but at most one would be a neutron decay because the neutron decay rate was quite low. Any others were background coincidences where the electron and proton events were uncorrelated in time. As a result, the background in the raw wishbone was completely flat and structureless, lacking the usual exponential shape of a random time spectrum, and background subtraction was relatively simple. Due to the longer time scale dead time effects observed in the NG-C data, we were unable to use the same method. Instead we kept only the earliest electron in the 11 $\mu$s wide coincidence time window of each proton and discarded any others. This change produced three important effects:
\begin{enumerate}
\item A random background coincidence could preempt a neutron decay event if the background electron event occurred earlier in time. This removed an estimated 20 \% of usable neutron decays from the data with a resulting loss of statistics. 
\item Neutron decay protons appear in a coincidence region of (3--4.5) $\mu$s after the beta electron as can be seen in figure \ref{F:wishRaw}. If an event appeared in this region, there could not have been an earlier electron event during the previous 5 $\mu$s, otherwise the coincidence region event would have been preempted. Note that earlier electrons correspond to longer proton TOF in the wishbone plot. This enforced the $>$4 $\mu$s dead time requirement described above.
\item The random background coincidences now have the usual exponential time structure. A more intricate method is needed to subtract background and correct for dead time.
\end{enumerate} 
\par
We begin with the assumption that the raw wishbone plot contains only neutron decay coincidence events and random background coincidences. This is reasonable because we do not expect physical correlations in background events in the time range (1--10) $\mu$s. The vast majority of background comes from gamma rays produced by neutron capture in the electrostatic mirror, collimator, and other nearby materials. The remainder is radioactive decay, guide hall background, and cosmic rays. Weak decays from neutron capture may produce correlations, but at much longer times. 
The others produce only prompt coincidences well within 1 $\mu$s. 
\par
Consider a vertical slice of the raw wishbone at a particular beta energy. Let the neutron decay wishbone function be bounded by proton TOF values $t_0$ and $t_1$. For $t<t_0$ the background has an exponential shape
\begin{equation}
\label{E:B0}
B(t < t_0) = c_0e^{Rt}
\end{equation}
and for $t>t_1$ a similar exponential shape
\begin{equation}
\label{E:B1}
B(t>t_1) = c_1e^{Rt}.
\end{equation}
Note that these are positive exponentials because larger $t$ (larger proton TOF) corresponds to earlier electron event time. The rate parameter $R$ is the same in both regions; it is the random background electron event rate at the energy of this wishbone slice. The constants $c_0$ and $c_1$ are different; their ratio $c_0/c_1 < 1$ is the probability that no neutron decay electron was detected, with proton TOF in the neutron decay window $t_0<t<t_1$, for a given proton event. The values of $R$, $c_0$, and $c_1$ are found by fitting the data simultaneously in the two regions outside the neutron decay window.
\par
Inside the neutron decay window the background shape is more complicated; at each point in $t$ it depends on the probability that a background electron was not preempted by a neutron decay electron prior to that point, {\em i.e.}
\begin{equation}
\label{E:B01}
B(t_0<t<t_1) = c(t)e^{Rt}
\end{equation}
with
\begin{equation}
\label{E:ct}
c(t) = c_1 - (c_1 - c_0) \frac{\int_t^{t_1} N(t') dt'}{\int_{t_0}^{t_1} N(t') dt'}.
\end{equation}
Here $N(t)$ is the neutron decay wishbone function that can be obtained by subtracting the background $B(t)$ from the measured spectrum and applying the dead time correction factor $e^{-R(t_1-t)}$. We start with an estimate for $N(t)$ and find the background $B(t)$ using equations \ref{E:B0}--\ref{E:ct}. Subtracting $B(t)$ from the measured wishbone slice yields an improved, measured result for $N(t)$ and we repeat the process iteratively until the resulting background subtracted wishbone function $N(t)$ is stable (typically three iterations). We note that $c_0/c_1 \approx 0.99$ in the beta energy range of interest (100 keV--400 keV) so the function $c(t)$ in equation \ref{E:ct} affects the background subtraction at the 1 \% level. This background subtraction algorithm was extensively tested using pseudodata and it worked very effectively.
\par
\begin{figure}
\centering
\includegraphics[width = 6in]{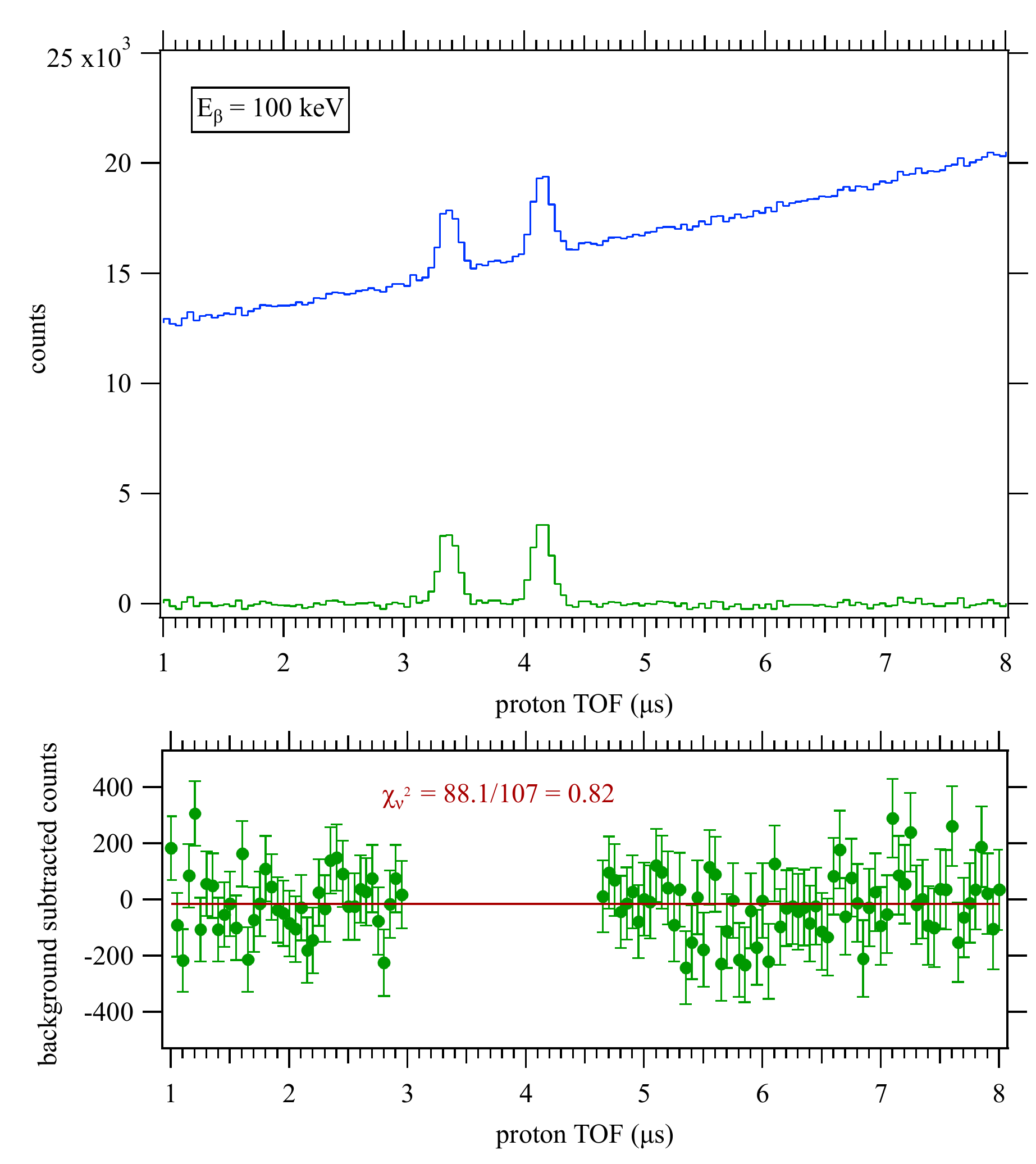}
\caption{\label{F:wishE100} Top: A 20-keV wide wishbone slice centered at beta energy 100 keV (blue), and the same wishbone slice after subtracting background (green). Bottom:
The same background subtracted slice fit to a horizontal line, excluding the neutron decay region (3--4.6 $\mu$s). Error bars are statistical.}
\end{figure}
\begin{figure}
\centering
\includegraphics[width = 6in]{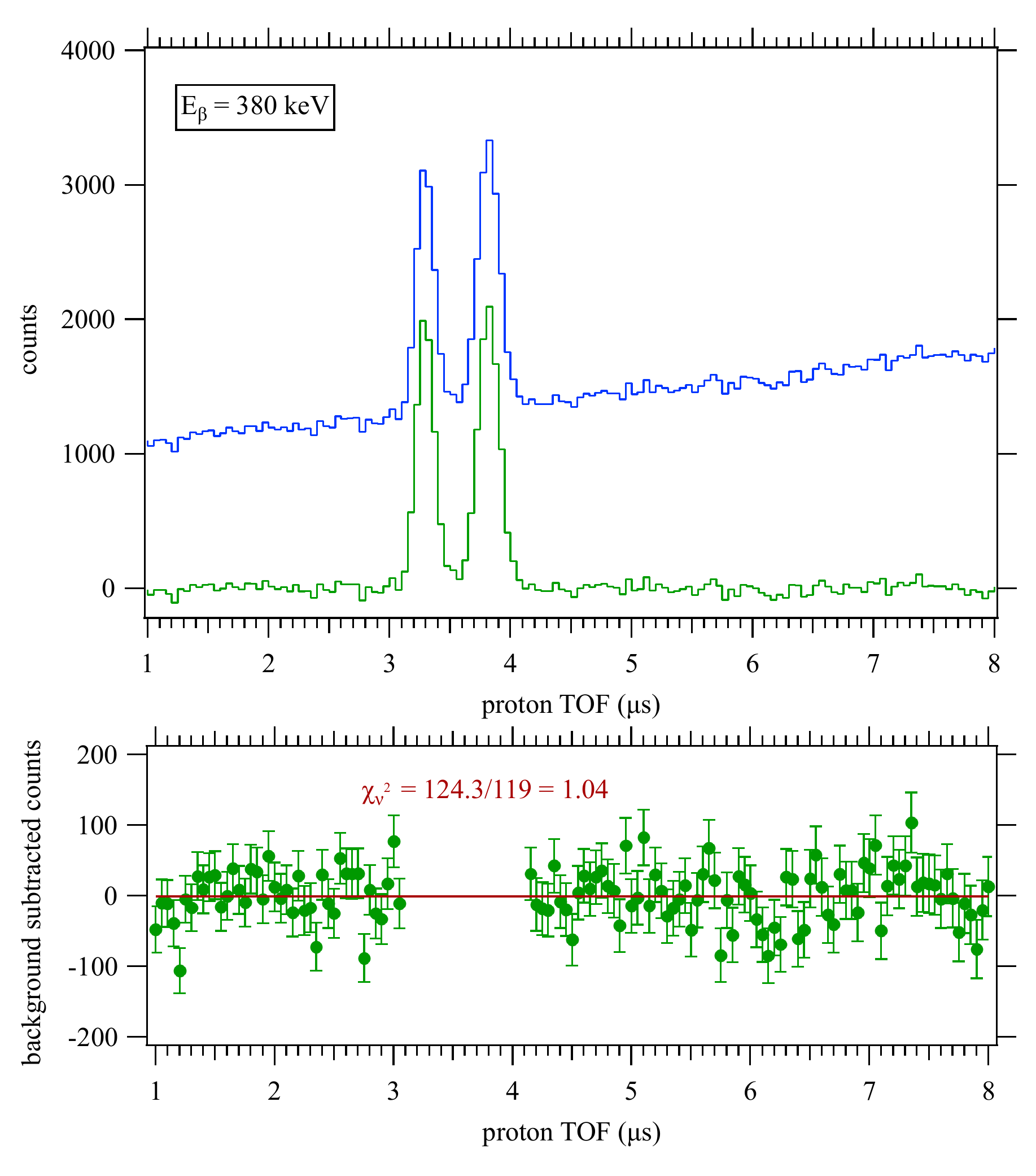}
\caption{\label{F:wishE380} Top: A 20-keV wide wishbone slice centered at beta energy 380 keV (blue), and the same wishbone slice after subtracting background (green). Bottom:
The same background subtracted slice fit to a horizontal line, excluding the neutron decay region (3.1--4.1 $\mu$s). Error bars are statistical.}
\end{figure}

Figure \ref{F:wishE100} shows a 20 keV wide vertical slice (blue) of the raw wishbone (figure \ref{F:wishRaw}), centered at 100 keV, the lowest beta energy that was used in the final analysis. Also shown is the same slice after background subtraction (green), {\em i.e.} the measured neutron decay wishbone function $N(t)$. The background outside the neutron decay window is flat and without apparent structure. The bottom plot in the figure is a fit of the same background-subtracted slice to a zero-slope line with the neutron decay window (3--4.6) $\mu$s excluded. The variation in counts is consistent with Poisson statistical fluctations. Figure \ref{F:wishE380} shows similar plots for a 20 keV wide vertical slice of the raw wishbone (figure \ref{F:wishRaw}) centered at 380 keV, the highest beta energy that was used in the final analysis. As can be seen here, the signal to background ratio (S/B) was strongly dependent on beta energy. In the energy range used in the analysis, $E_e$ = 100 keV--380 keV, the average S/B was 0.2.
\par
\begin{figure}
\centering
\includegraphics[width = 5in]{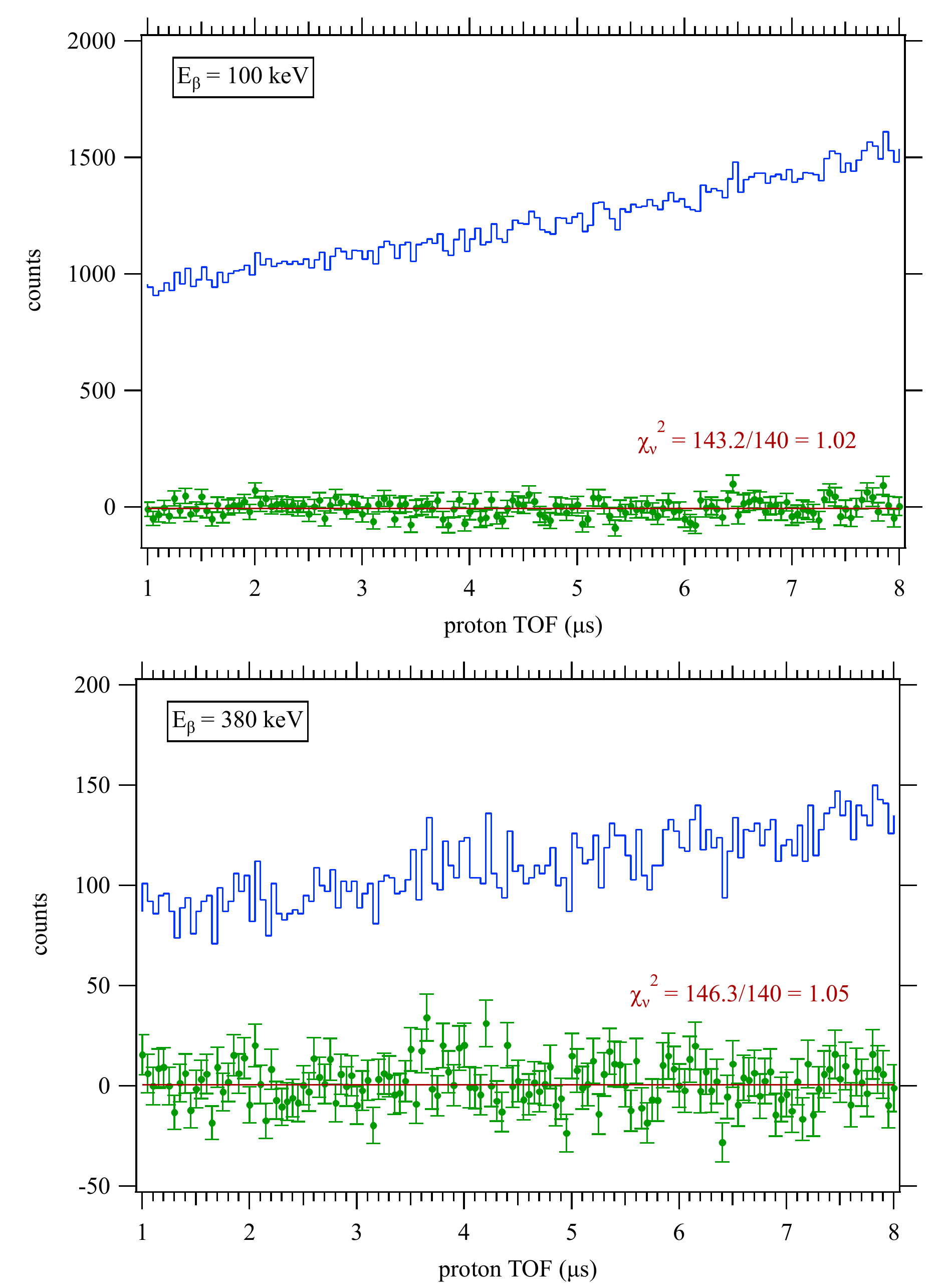}
\caption{\label{F:revEM} 20-keV wide wishbone slices centered at 100 and 380 keV for a data series where the polarity of the electrostatic mirror was reversed, so neutron decay protons could not be detected.
The upper curve (blue) is the raw wishbone and the lower curve (green) is after subtracting background and fitting to a horizontal line. Error bars are statistical.}
\end{figure}
\begin{figure}
\centering
\includegraphics[width = 6in]{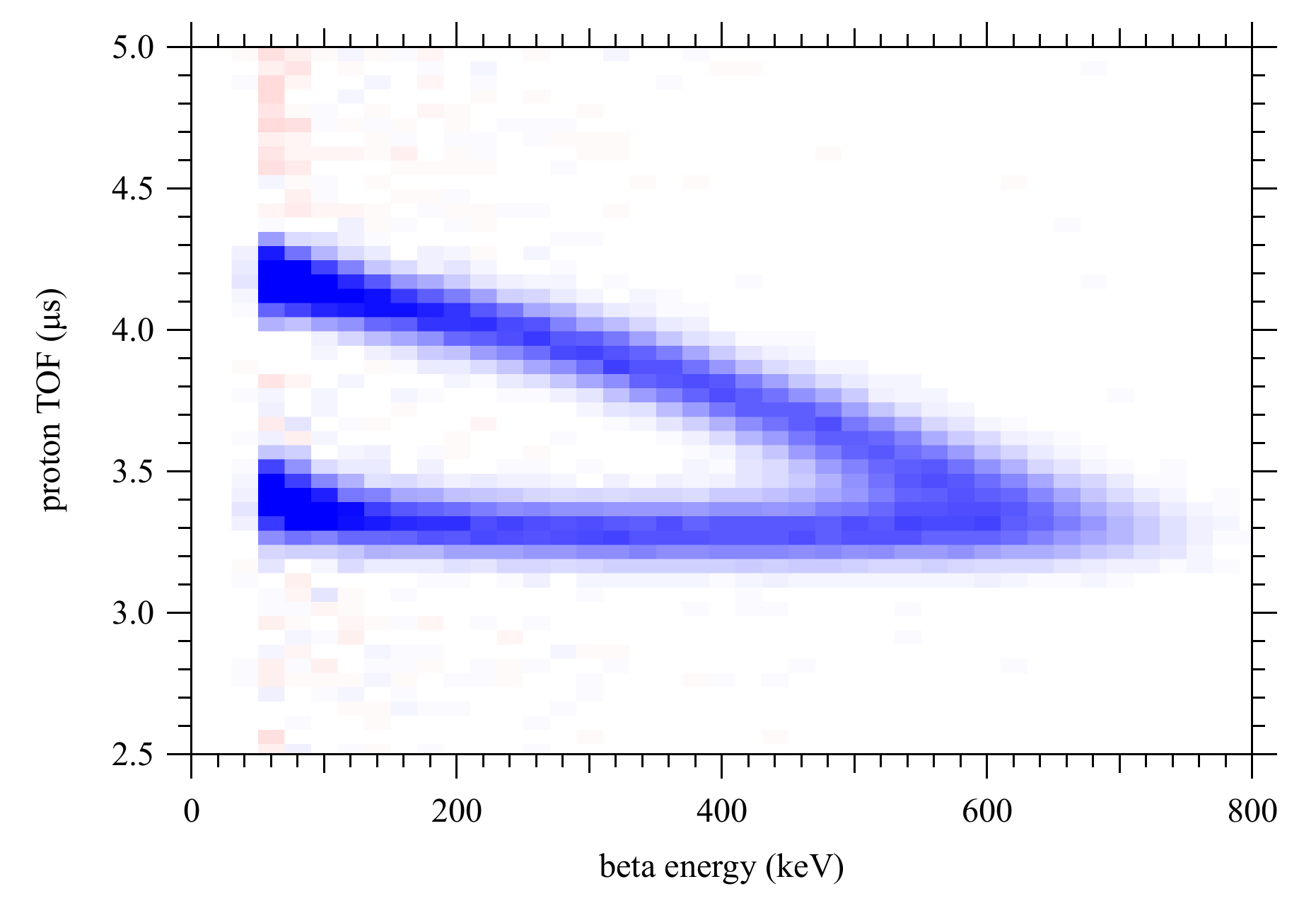}
\caption{\label{F:wishboneES} A background-subtracted wishbone plot (data from figure \ref{F:wishRaw}). Blue points are positive and red are negative.}
\end{figure}

During the experimental run, as a systematic check, we collected 19 hours of beam data with the polarity of the electrostatic mirror reversed. This prevented all neutron decay protons from reaching the proton detector with minimal effect on background coincidences. Data from this run are shown in figure \ref{F:revEM}, again 20 keV wide slices centered at beta energies 100 keV and 380 keV. Other than the expected exponential there is no apparent structure in the background inside or outside the neutron decay window. The green points are after background subtraction using the same algorithm as for the neutron decay data described above, and fitting to a zero-slope line. A full background-subtracted and deadtime-corrected wishbone plot is shown in figure \ref{F:wishboneES}, obtained from the data shown in figure \ref{F:wishRaw}. Blue points are positive and red points are negative (due to background subtraction). 

\subsection{Energy Calibration Fit}
\label{SS:eCal}
\begin{figure}
\centering
\includegraphics[width = 6in]{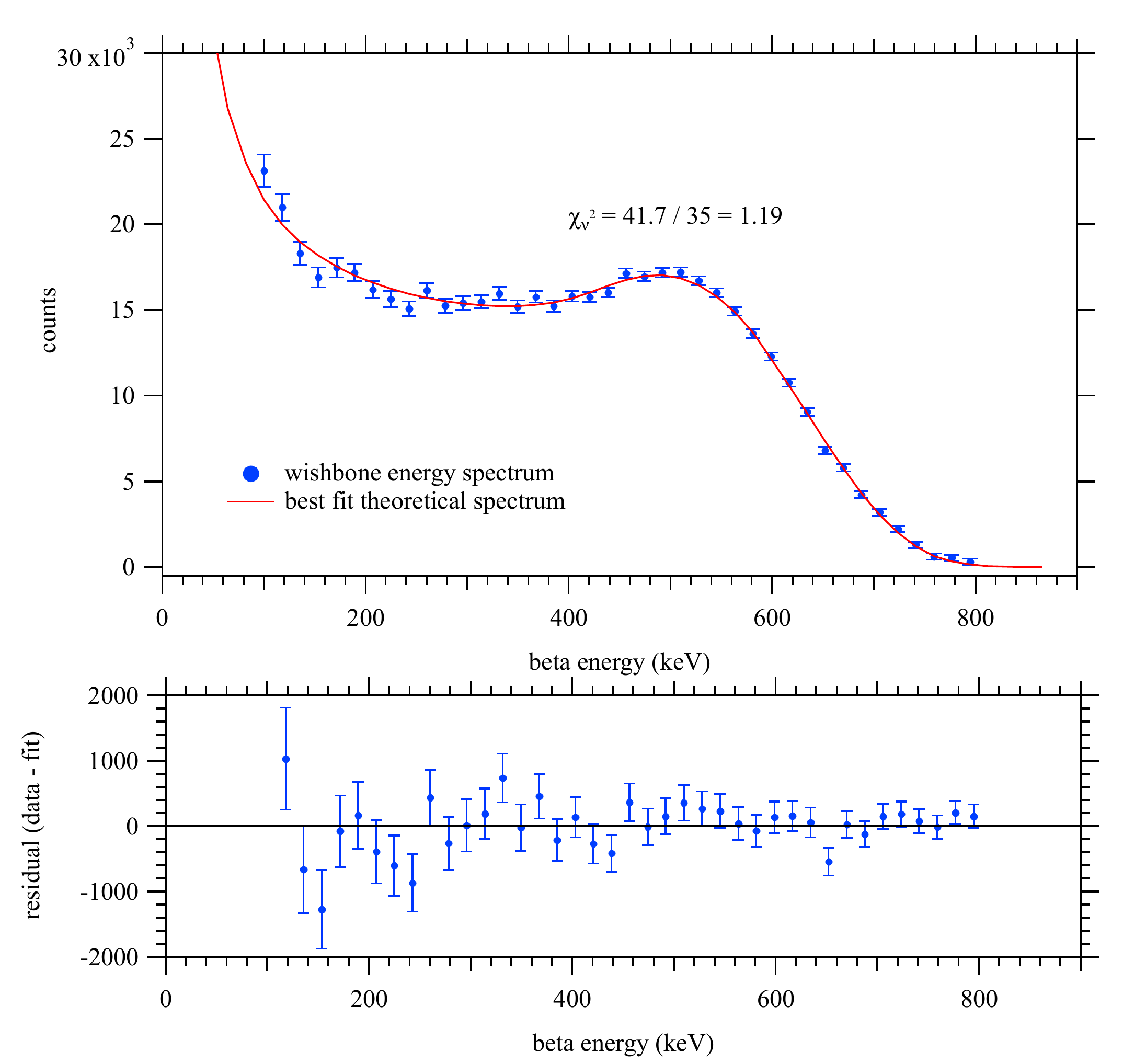}
\caption{\label{F:spectFit} Top: The wishbone energy spectrum, {\em i.e.} the background subtracted wishbone (figure \ref{F:wishboneES}) summed over proton TOF and the best fit theoretical spectrum. Bottom: Fit residuals (data minus fit). Error bars are statistical.}
\end{figure}

The absolute beta energy calibration was monitored during the run by collecting data from {\em in situ} conversion electron sources ($^{113}$Sn and $^{207}$Bi) 3--4 times per week, interleaved with the neutron decay data series.   A more robust and precise energy calibration was obtained later for each data group using the neutron decay beta spectrum. Figure \ref{F:spectFit} (top) shows the wishbone energy spectrum, which is the background subtracted wishbone data (figure \ref{F:wishboneES}) summed over proton TOF. The corresponding theoretical spectrum is the Fermi beta energy spectrum $F(E_e) = E_e |{\bf p_e}| (Q-E_e)^2$ found in equation \ref{E:JTWeqn}, modified by the constraints on beta and proton momenta for coincidence events imposed by the aCORN collimation. The solid curve in figure \ref{F:spectFit} (top) is this theoretical spectrum computed numerically using the collimator diameters, axial magnetic field strength, and neutron beam geometry. The theoretical function was fit to the data to minimize chi-squared, with four variable free parameters:
\begin{itemize}
\item An overall multiplicative scale factor
\item A linear energy calibration slope
\item A linear energy calibration offset
\item The theoretical function was convoluted with a normalized Gaussian energy response function $G(E,E') = \frac{1}{\sqrt{\pi CE'}}\exp(-(E-E')^2 / CE')$, based on the expected $\sqrt{E}$ resolution-width dependence of the scintillator detector. The constant $C$ was a free parameter in the fit.
\end{itemize}
Acceptable fits were obtained as illustrated in figure \ref{F:spectFit}. With this method the wishbone data were self-calibrating for beta energy. This result also supports the success of the background subtraction, the absence of extraneous structure in the data, and the effectiveness of the backscatter suppression which obviated the need for a low energy tail in the beta response function.  We note that the wishbone energy spectrum in figure \ref{F:spectFit} is insensitive to the wishbone asymmetry and the value of the $a$-coefficient so these fits had no  bearing on the asymmetry analysis (section \ref{SS:asymUp}), other than to provide the absolute beta energy scale.

\subsection{Wishbone Asymmetry Analysis, Magnetic Field Up}
\label{SS:asymUp}
\begin{figure}
\centering
\includegraphics[width = 6in]{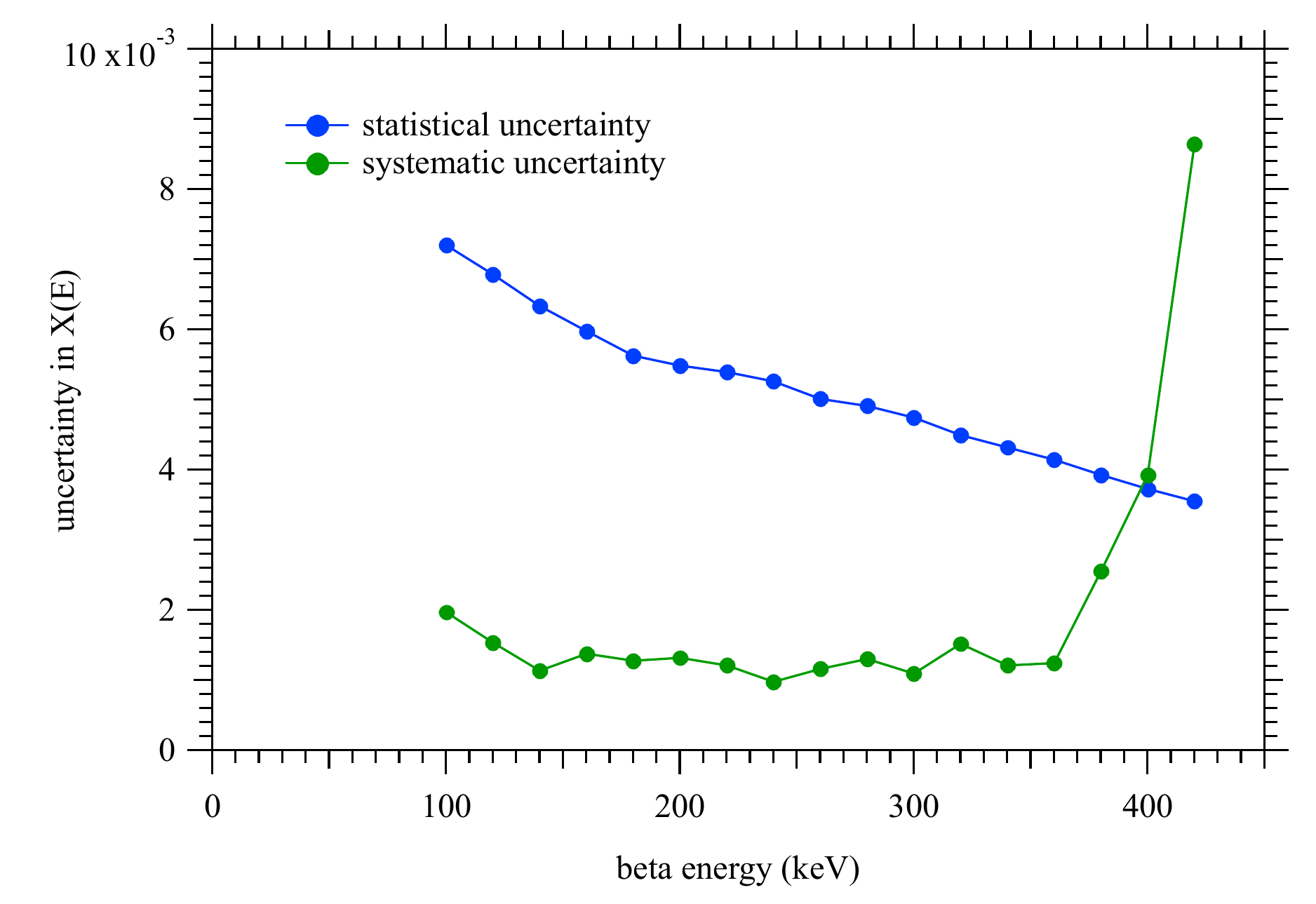}
\caption{\label{F:wbAsymSys} The estimated systematic uncertainty in computing the wishbone asymmetry $X(E)$ from the data, compared to the Poisson statistical uncertainty.}
\end{figure}

To calculate the wishbone asymmetry $X(E)$ for data taken with the magnetic field up direction (B$_{\rm up}$), we start with 20-keV wide vertical slices of the background-subtracted wishbone plot (figure \ref{F:wishboneES}) for each data group. The background-subtracted histograms (green) in figures \ref{F:wishE100}, \ref{F:wishE380} are examples of these. From equation \ref{E:asym} we have
\begin{equation}
X(E) = \frac{N^I(E) - N^{II}(E)}{N^I(E) + N^{II}(E)}
\end{equation}
where $N^I(E)$ and $N^{II}(E)$ are the counts in the fast (left) and slow (right) peaks, respectively, for each energy slice, summed over all B$_{\rm up}$ data groups. Because the fast and slow peaks tend to overlap a bit, we are faced with the questions of which TOF bin to use to separate them, and how to apportion the counts within that bin. For this we use Monte Carlo data as a guide. We take a high-statistics Monte Carlo wishbone slice for each beta energy, and find the TOF bin and its apportionment that reproduces the exactly correct wishbone asymmetry based on the input value of the $a$-coefficient. This can always be done in spite of the slight overlap of the fast and slow peaks. We expect a systematic uncertainty in this procedure that will be small for low beta energy where the overlap is negligible, and large for beta energy above $\approx$400 keV where the overlap becomes significant. To estimate this uncertainty, we assume that the correct apportionment of the TOF separation bin lies somewhere between 100 \% of its counts to the slow peak and 100 \% to the fast peak, and assign this full range a 95 \% C.L. ($\pm\, 2 \sigma$). It then follows that the $1\sigma$ systematic uncertainty equals one-half the counts in the separation bin divided by the total counts in the fast and slow wishbone peaks. Figure \ref{F:wbAsymSys} shows the average systematic uncertainty in the wishbone asymmetry using this prescription, compared to the Poisson statistical uncertainty in $X(E)$ for all B$_{\rm up}$ data. We restrict the $a$-coefficient analysis to the energy range where this systematic uncertainty is less than the statistical, {\em i.e.} up to 380 keV.
\par
\begin{figure}
\centering
\includegraphics[width = 6in]{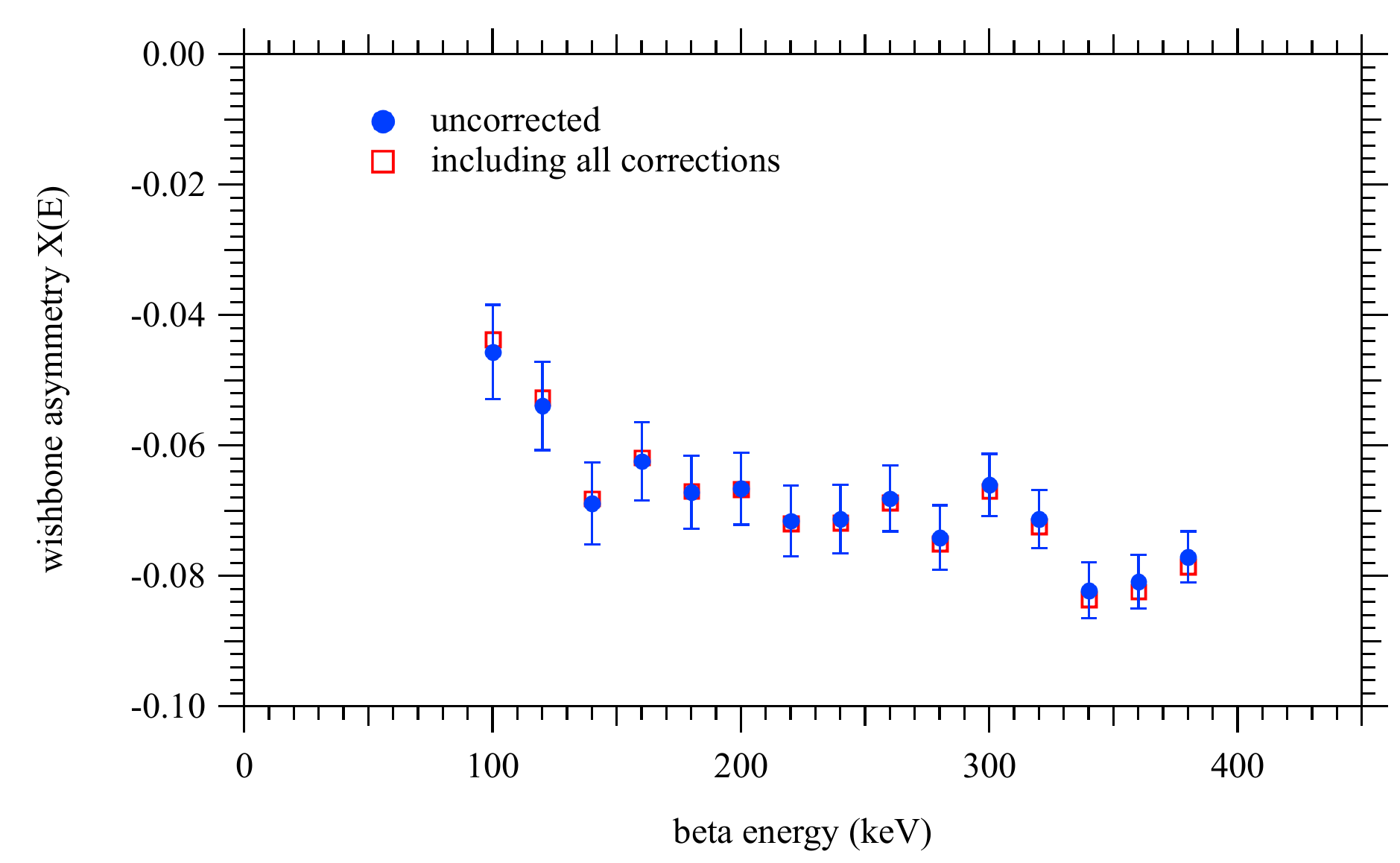}
\caption{\label{F:wbAsymUp} The wishbone asymmetry $X(E)$ for the combined $B_{\rm up}$ data, uncorrected with statistical error bars, and with all corrections. }
\end{figure}

The wishbone data for beta energy $\leq$80 keV has a number of issues:
\begin{itemize}
\item Beta electrons may have zero axial momentum and still satisfy the transverse momentum acceptance, adding a tail to the wishbone TOF.
\item Some beta electrons will miss the active region of the beta spectrometer, as shown by Monte Carlo simulation, complicating the geometric function $f_a(E)$.
\item The wishbone signal/background is very poor in this energy region and the background subtraction is imperfect.
\end{itemize}
From the above considerations we choose the energy range 100 keV--380 keV for the $a$-coefficient analysis. The uncorrected wishbone asymmetry $X(E)$ for all B$_{\rm up}$ data is shown in figure \ref{F:wbAsymUp}.

\subsection{Systematic Effects and Corrections}
\label{SS:Systematics}
\subsubsection{Electrostatic Mirror}
\label{SSS:ESmirror}
\begin{figure}
\centering
\includegraphics[width = 6in]{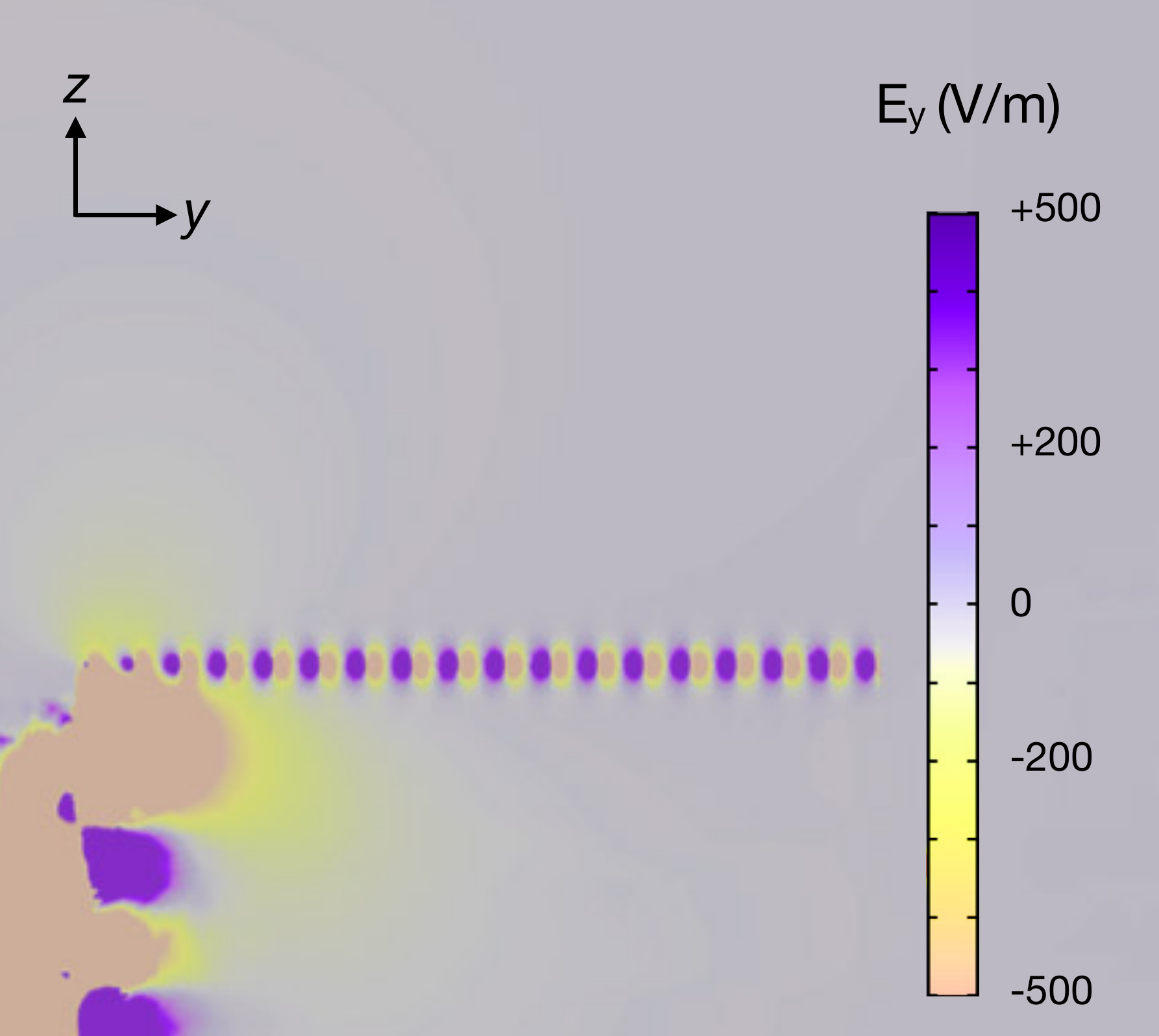}
\caption{\label{F:Mirror7} A COMSOL finite element map of the transverse electric field inside the NG-C electrostatic mirror, in the region near the upper wire grid through which the protons pass.}
\end{figure}

\begin{figure}
\centering
\includegraphics[width = 6in]{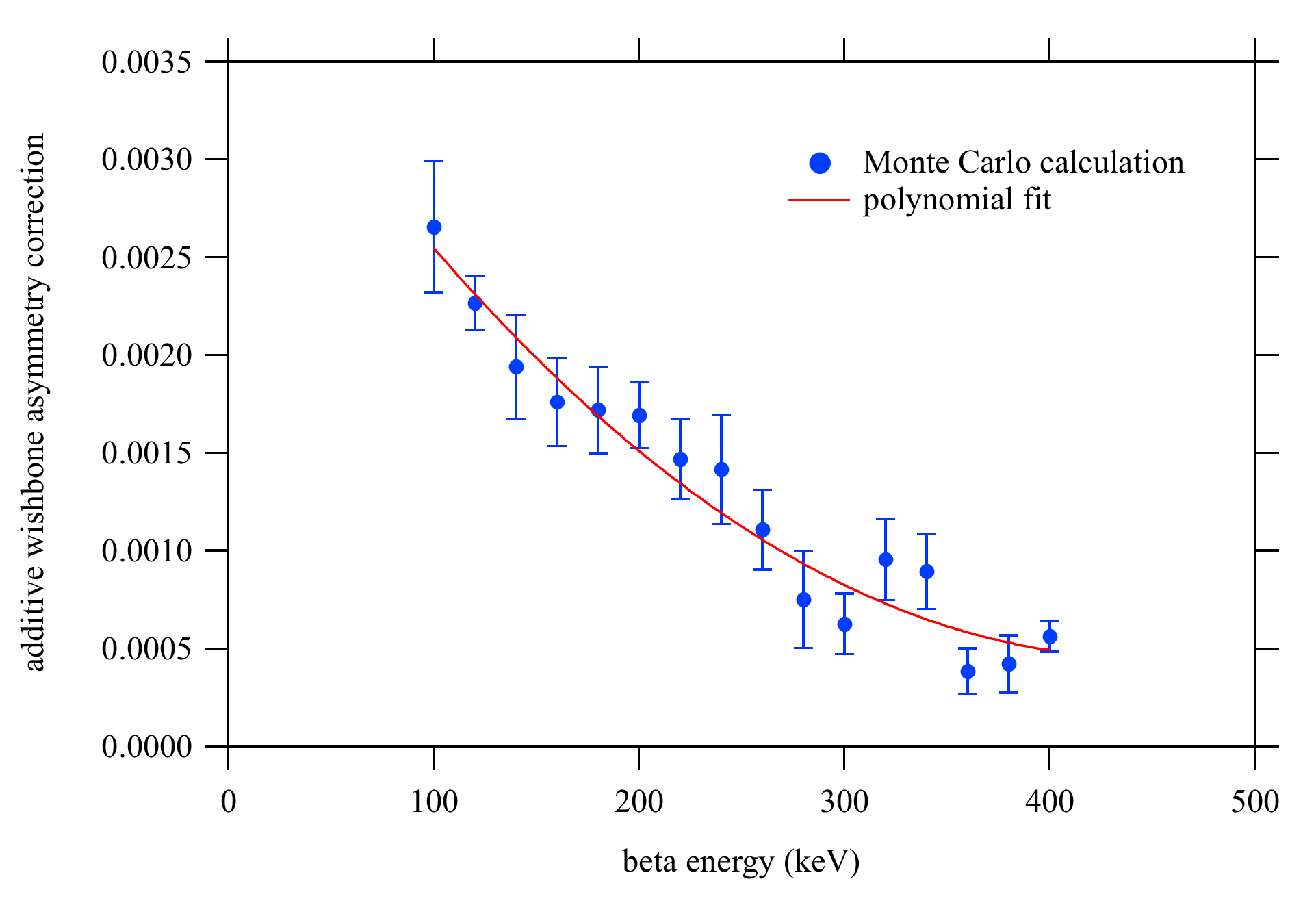}
\caption{\label{F:EScorrect} The electrostatic mirror correction calculated by proton transport Monte Carlo using the 3D COMSOL model of the electric field. The red curve is a smoothed
average obtained by fitting the Monte Carlo data to a second order polynomial. Error bars are statistical.}
\end{figure}

The electrostatic mirror was designed to provide an approximately uniform axial electric field in the proton transport region. Protons associated with group I and II wishbone events tend to have different trajectories inside the mirror so the presence of transverse electric fields will cause a bias in their transmission within the proton collimator. Through Monte Carlo studies we found that a 0.1 \% uniform transverse electric field, relative to the axial, produces a 0.5 \% false wishbone asymmetry. Due to the precision of its construction and alignment (see figure \ref{F:insertAlign}) the uniform transverse field was much smaller than this. However it is unfortunately not possible to avoid significant transverse electric fields in the vicinity of the upper (grounded) wire grid. In the NG-6 run this effect gave the largest correction to the result: ($5.2 \pm 1.1$) \% \cite{Dar17}. For the NG-C run the grid support structure was modified and the upper linear grid was replaced with the crossed wire grid shown in figure \ref{F:newGrid}. A detailed 3D COMSOL \cite{COM} model, depicted in figure \ref{F:Mirror7}, was built to calculate the resulting electric field shape. This field map was input to the aCORN proton transport Monte Carlo to calculate the beta-energy dependent correction shown in figure \ref{F:EScorrect}, an overall relative correction to the wishbone asymmetry of ($1.36 \pm 0.27$) \%. The uncertainty was calculated using a standard 20 \% relative uncertainty that we chose and assigned to all Monte Carlo corrections in this experiment. We regard this uncertainty to be an overestimate as the electric field and proton transport calculations are expected to be much more accurate than 20 \%.

\subsubsection{Magnetic Field}
\label{SSS:magfield}
The proton collimation is affected by both the shape and absolute value of the magnetic field in the proton transport region. In particular, a transverse magnetic field will cause a bias in proton collimation and thence a false wishbone asymmetry. For a radially symmetric transverse field the effect averages out. The absolute value of the magnetic field is used to calculate the geometric function $f_a(E)$; an error in the absolute field will result in a proportional error in the $a$-coefficient result. 
\par
Through Monte Carlo analysis we have found that the false asymmetry is proportional to the average magnitude of the transverse magnetic field in the proton transport region. An average of 4 $\mu$T produces a wishbone asymmetry of $\Delta X = -3.4 \times 10^{-4}$ which is about 0.5 \% of the $a$-coefficient asymmetry. Based on the field maps, the average transverse field in the B$_{\rm up}$ configuration was 1 $\mu$T giving a systematic error in the asymmetry of $\Delta X = -8.5 \times 10^{-5}$ and we assign an uncertainty equal to the size of the correction.
\par
The absolute axial magnetic field was determined from NMR measurements on a glass cell filled with spin-polarized $^3$He that was lowered into the proton collimator from above. For B$_{\rm up}$  the result was 
B$_{\rm axial}$ = 36.39(11) mT, which leads to an uncertainty of 0.3 \% in the calculated $f_a(E)$.

\subsubsection{Electron Backscatter}
\label{SSS:backscatter}
Approximately 5 \% of electrons that strike the active energy detector will backscatter from it and the energy deposited is incomplete, producing a low energy tail in the electron response function. Such backscattered events have two undesirable effects: 1) they tend to fill in the gap between the wishbone branches (see figure \ref{F:MCwishbone}) and confound our ability to cleanly separate group I and group II events; and 2) they systematically shift events from group II into group I, causing a false positive wishbone asymmetry. The backscatter veto system in the beta spectrometer was used to mitigate this problem. Electrons may also scatter from the beta collimator with similar effect. These cannot be vetoed, but the collimator was designed to limit the probability of a scattered electron to reach the beta spectrometer to 0.3 \%, as verified in a PENELOPE simulation. Electron scatter from other materials or residual gas, and electron Bremsstrahlung, were investigated during the NG-6 run and found to be negligible \cite{Col17}.
\par
\begin{figure}
\centering
\includegraphics[width = 6in]{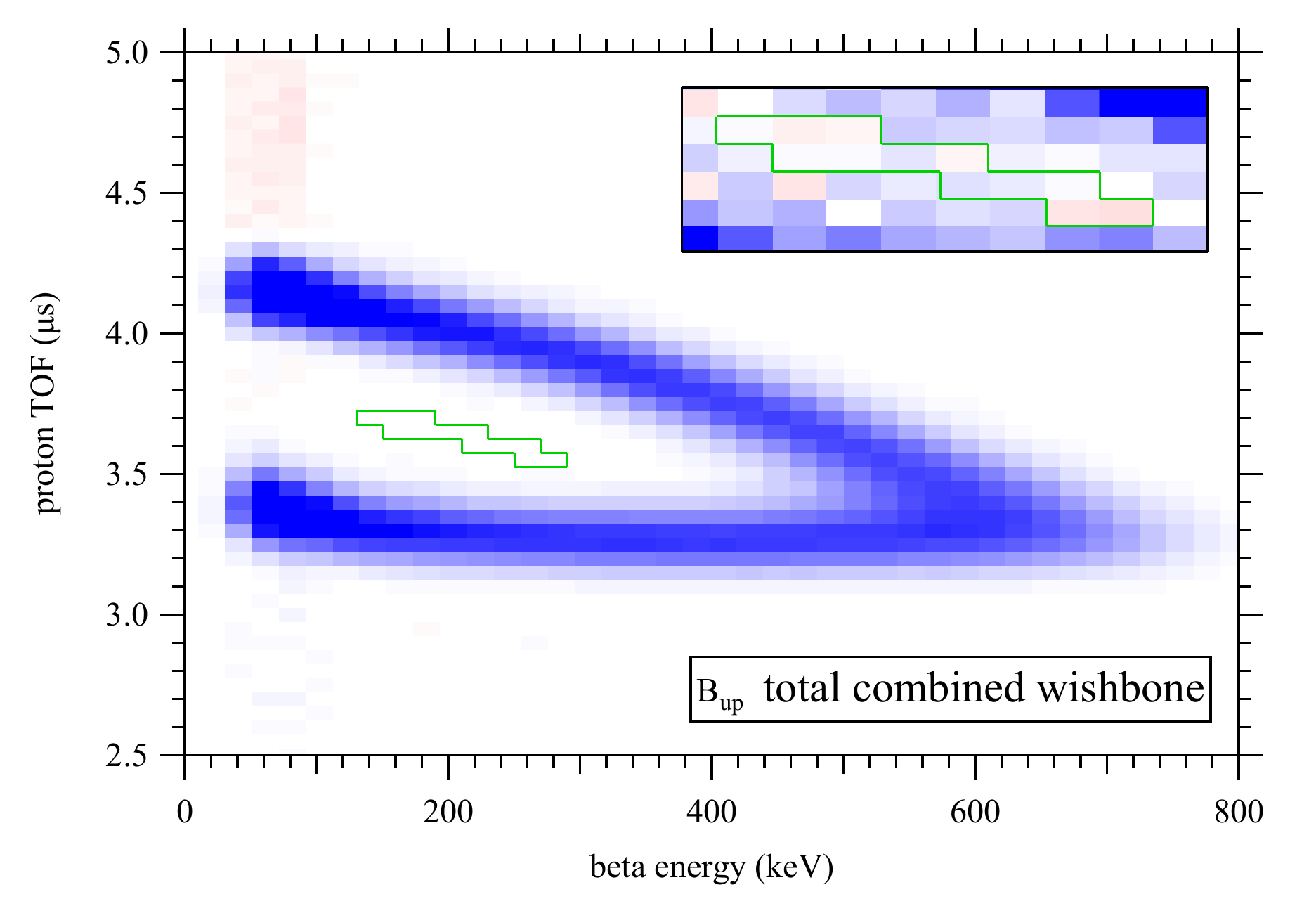}
\caption{\label{F:eScatterWB} A combined background-subtracted wishbone plot with all B$_{\rm up}$ data. The event total in the region outlined in green was used to test for the presence of a low energy tail in the detected electron response function due to electron scattering. The inset shows the same green-outlined region with an expanded color scale. Blue points are positive counts, red points are negative due to the background subtraction.}
\end{figure}

Our best test for electron backscatter effects was in the wishbone data. We looked for an excess of events in the gap between the wishbone branches and compared to a Monte Carlo wishbone that included a low energy scattering tail. Figure \ref{F:eScatterWB} shows a combined background-subtracted wishbone plot with all B$_{\rm up}$ data. The choice of the gap region, indicated in green, required some optimization. We want to use a large region while avoiding the tails of the wishbone branches and avoiding low energies where the background subtraction uncertainty is large.  A nonzero total of counts in this region can be attributed to non-vetoed backscattered electrons but would also include contributions from electron collimator scattering, electron scattering from the wire grid (section \ref{SSS:gridLoss}), and proton collimator scattering (section \ref{SSS:pScatter}). The number of counts in the chosen gap is 62 $\pm$ 490, consistent with zero. The uncertainty is due to the background subtraction. We take the total $62+490 = 552$ counts to be the $1\sigma$ upper limit due to non-vetoed backscattered electrons, and zero to be the lower limit. We generated Monte Carlo wishbone data, including a flat tail in the electron energy response function, and varied the tail area to achieve a count rate in the gap region that equals the $1\sigma$ upper limit. The resulting tail area was 0.59 \% of the peak, which produces an average false asymmetry of +1.5 \%. Therefore our systematic error due to electron backscatter is (+$0.75\pm0.75$) \%.

\subsubsection{Electron Energy Loss in Grid}
\label{SSS:gridLoss}
Beta electrons pass through the positive grid at the bottom of the electrostatic mirror. The grid is composed of parallel wires, diameter 100 $\mu$m, made of 2 \% beryllium copper with approximately 1 $\mu$m coatings of nickel and gold. The wire spacing is 2 mm, so the geometric probability of striking a grid wire is approximately 5 \%. When an electron strikes a wire the main systematic effect comes from energy loss. A beta electron will generally pass through the wire and lose typically about 100 keV. Electrons may also be scattered into a different direction, but to first order the probability of scattering into the collimator acceptance is the same as the probability of scattering out of it, and because the wishbone asymmetry is insensitive to beta collimation this does not create a systematic error. Energy loss in a grid wire is similar to backscatter from the beta spectrometer in its effect, but instead of producing a broad low energy tail it produces a small low energy shoulder on the energy response function, which is less of a problem. Energy loss in the grid and its effect on the electron energy response was calculated using the NIST ESTAR data base \cite{Estar}. The associated error in the wishbone asymmetry is (+1.0$\pm$0.2) \%, using our 20 \% standard Monte Carlo uncertainty.

\subsubsection{Beta Energy Calibration}
As discussed in section \ref{SS:eCal}, the most precise beta energy calibration comes from a fit to the wishbone data. The combined calibration from all B$_{\rm up}$ data gives an overall energy uncertainty of 
$\sigma(E) = \pm$0.48 \%. The corresponding uncertainty in the wishbone asymmetry is
\begin{equation}
\sigma(X) = a \frac{\partial f_a(E)}{\partial E}  \sigma(E)
\end{equation}
which has an average value of 0.27 \% in the energy range 100 keV--380 keV.

\subsubsection{Proton Energy Threshold}
\begin{figure}
\centering
\includegraphics[width = 6in]{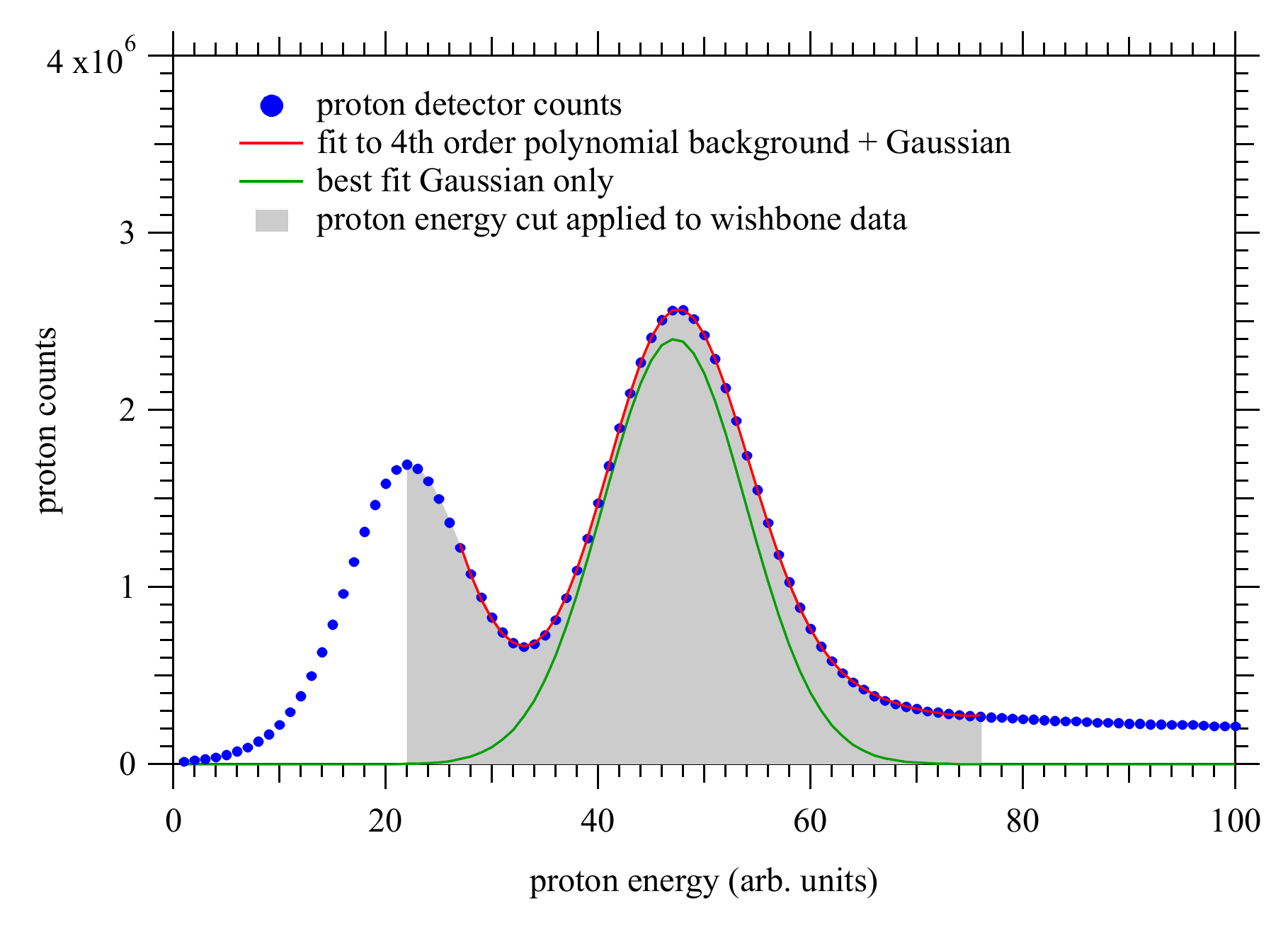}
\caption{\label{F:pThresh} A typical proton energy spectrum (blue) fit to a 4$^{\rm th}$ order polynomial background function plus a Gaussian (red). The resulting Gaussian alone is shown in green. The soft energy threshold of the PIXIE-16 takes effect below channel 27.  The slight loss of protons below threshold  has a negligible effect on the wishbone asymmetry.}
\end{figure}

Protons associated with group I and II coincidence events differ in kinetic energy by an average of 380 eV. Both groups of protons are preaccelerated by the electrostatic mirror and then, after passing through the proton collimator, accelerated to a final energy of about 30 keV by the proton focusing electrodes and detector. While this difference in energy is a small fraction of the detected energy, protons near threshold nevertheless contain a slightly higher fraction of group II protons. If these are not completely counted, a false negative wishbone asymmetry results. In the NG-6 aCORN run about 1.2 \% of protons were excluded by the PIXIE threshold which lead to a  3.0 \% false asymmetry \cite{Dar17}. For the NG-C run we significantly lowered the PIXIE energy threshold (see section \ref{SS:daqNGC}). Figure \ref{F:pThresh} shows a fit of a typical proton energy spectrum fit to a Gaussian plus a 4$^{\rm th}$ order polynomial background function to extract the Gaussian component. The fraction of events excluded by the threshold is less than 0.02 \% and the resulting false asymmetry is negligible.
\subsubsection{Collimator Insert Alignment}
\begin{figure}
\centering
\includegraphics[width = 5in]{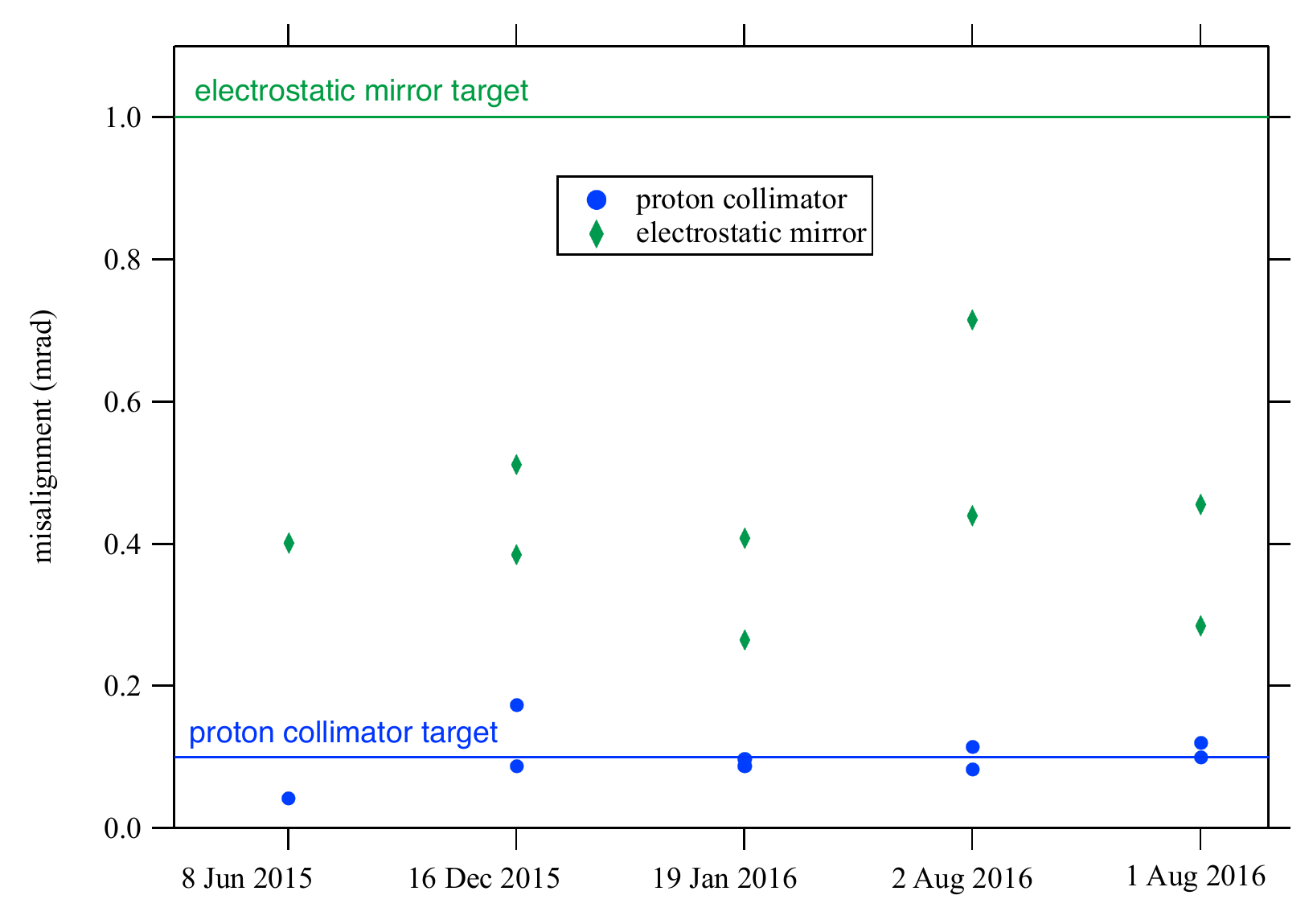}
\caption{\label{F:insertAlign} A summary of optical insert alignment checks made over the course of the run. Multiple points are independent measurements made by two people.}
\end{figure}

A small angular misalignment $\phi_{\rm coll}$ (radians) of the proton collimator is equivalent to a uniform transverse magnetic field $B_{\rm trans} = \phi_{\rm coll} B_{\rm axial}$. Figure \ref{F:insertAlign} shows a summary of the collimator alignment measurements. The variation in results obtained by two independent observers for the same misalignment strongly suggests that the overall variation is due mostly to measurement error rather than differences in the actual misalignment. Therefore we take the mean misalignment and the standard deviation (square root of variance) from all nine measurements: $\phi_{\rm coll} = (0.101 \pm 0.035)$ mrad. Using B$_{\rm axial}$ = 0.0364 T we have $B_{\rm trans}$ = ($3.7\pm1.3$) $\mu$T. Using the Monte Carlo result described in section \ref{SSS:magfield}, this results in $\Delta X = (-3.1 \pm 1.3) \times 10^{-4}$, where the standard 20 \% Monte Carlo uncertainty has been included in quadrature. Note that this effective transverse magnetic field is independent of that measured by the field mapper as the collimator was not present when the maps were made. Therefore we treat the collimator misalignment as an independent source of error.
\par
Similarly, a misalignment of the electrostatic mirror would introduce an approximately uniform transverse electric field. From Monte Carlo analysis we found that a 1 mrad misalignment will produce a false wishbone asymmetry of 
$\Delta X = -4 \times 10^{-4}$. The mean and standard deviation of the measured values shown in figure \ref{F:insertAlign} is $\phi_{\rm mirror} = (0.43 \pm 0.13)$ mrad corresponding to $\Delta X = -1.7 \pm 0.6 \times 10^{-4}$.
\subsubsection{Residual Gas Interactions}
Protons travel about 2 m from the decay region to the detector. If a proton interacts with residual gas during this trip it may be neutralized or scattered. Neutralized protons cause neutron decay events to be eliminated and they may introduce a false wishbone asymmetry due to the slight velocity-dependence of the neutralization probability. Scattered protons result in a larger TOF in the wishbone plot which may also result in a false wishbone asymmetry. Monte Carlo analyses showed that proton scattering and neutralization have opposite-sign effects on the asymmetry, and that their relative probability depends on the gas species. We accounted for this effect by collecting data for 134 hours with a deliberately higher pressure in the chamber, effected by partially closing a gate valve to the turbopump. The average pressure in the proton collimator during the high pressure run was $1.79 \times 10^{-3}$ Pa ($1.34 \times 10^{-5}$ torr), compared to the normal pressure of $8.0 \times 10^{-5}$ Pa ($6.0 \times 10^{-7}$ torr), a factor of 22 higher. Residual gas analyzer (RGA) measurements indicated that the gas was dominated by hydrogen and water (due to outgassing from the beta spectrometer plastic scintillator) at both pressures. 
\par
Comparing the wishbone asymmetry from the high pressure run, from beta energy 100 keV--380 keV, to that of the production B$_{\rm up}$ data, we found an average difference $\Delta X = -0.0024\pm0.0070$, consistent with no effect. We therefore estimate the systematic uncertainty due to residual gas interaction as $\sigma_X = 0.0070/22 = 3.2 \times 10^{-4}$.

\subsubsection{Proton Scattering from the Collimator}
\label{SSS:pScatter}
\begin{figure}
\centering
\includegraphics[width = 5in]{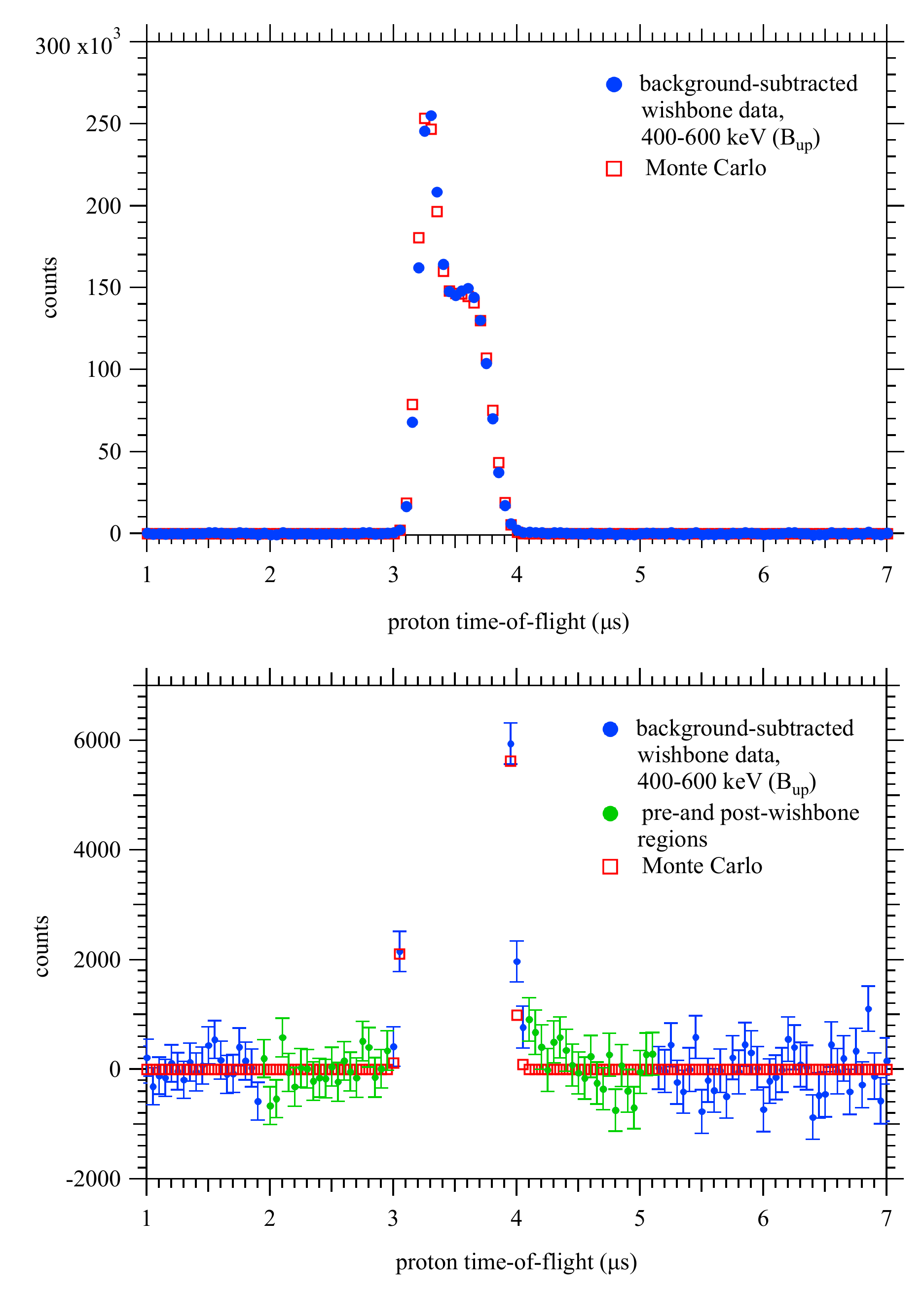}
\caption{\label{F:pScatter} Top: The total B$_{\rm up}$ wishbone proton TOF spectrum summed from 400--600 keV (blue) compared to the equivalent Monte Carlo proton TOF spectrum (red). Bottom: The same plot
with an expanded vertical scale and statistical error bars. The 1 $\mu$s wide regions pre- and post-wishbone used to estimate the proton scattering tail are shown in green.}
\end{figure}

A large number of neutron decay protons strike the aluminum knife edge elements of the proton collimator. A SRIM Monte Carlo study showed that for protons with energy in the range 2--3 keV, about 90 \% of those will be absorbed in the aluminum, 9.5 \% will emerge as neutral hydrogen atoms, and the remaining 0.5 \% emerge as bare protons, having lost an average of 2/3 of their kinetic energy. Many of those will subsequently strike the collimator again and be removed but some fraction will be detected with TOF that is systematically too large. Absorbed and neutralized protons are not detected and cause no systematic effect. Because protons are accelerated by the electrostatic mirror they have a minimum possible axial momentum while in the collimator. This sets an upper limit on the TOF for unscattered protons in the wishbone plot. Scattered neutron decay protons would appear beyond this maximum as a broad tail several $\mu$s in width, and we can study this effect in the wishbone plot. This effect is insensitive to beta energy, so it is useful to look at relative high beta energy where the statistical uncertainty due to background subtraction is smaller.  We use the beta energy range 400 keV--600 keV. Figure \ref{F:pScatter} (top) shows the total B$_{\rm up}$ wishbone proton TOF spectrum summed from 400 keV--600 keV compared to the equivalent Monte Carlo proton TOF spectrum. Figure \ref{F:pScatter} (bottom) is the same with an expanded vertical scale. We choose 1-$\mu$s wide regions just before and after the wishbone TOF peak where the Monte Carlo counts are zero and take the difference of their sums, post-wishbone minus pre-wishbone, which is $2296 \pm 2400$ counts. As a fraction of the wishbone peak area this is $0.0010 \pm 0.0011$, consistent with the SRIM estimate but also statistically consistent with zero. Comparing this to a Monte Carlo analysis where a proton scattering TOF tail was included, this corresponds to a systematic error in the wishbone asymmetry of $\Delta X = -0.00036\pm0.00038$.

\subsubsection{Proton Focusing}
The proton detector focusing system was designed to focus all neutron decay protons that were accepted by the proton collimator onto the active region of the surface barrier detector. The focusing efficiency, while very good, was not perfect. A small fraction of protons may strike the focusing electrodes or an inactive region of the detector, or miss the detector entirely. Because the average kinetic energies of the fast (group I) and slow (group II) protons differ slightly at the exit of the collimator, and the focusing efficiency is expected to depend on kinetic energy, imperfect proton focusing will lead to a systematic error in the wishbone asymmetry. This effect was studied computationally and experimentally.
\par
A simulation of the focusing assembly and related apparatus was developed using the software suite AMaze by Field Precision \cite{AMaze}. The relative positions of the surface barrier proton detector and ring and fork electrodes were accurately measured using a FARO coordinate measuring device \cite{FARO}. An auxiliary simulation produced neutron decay protons in the decay region and transported them to the exit of the proton collimator. These proton momenta were then fed into the AMaze simulation to track them to the detector.
\par
\begin{figure}
\centering
\includegraphics[width = 6in]{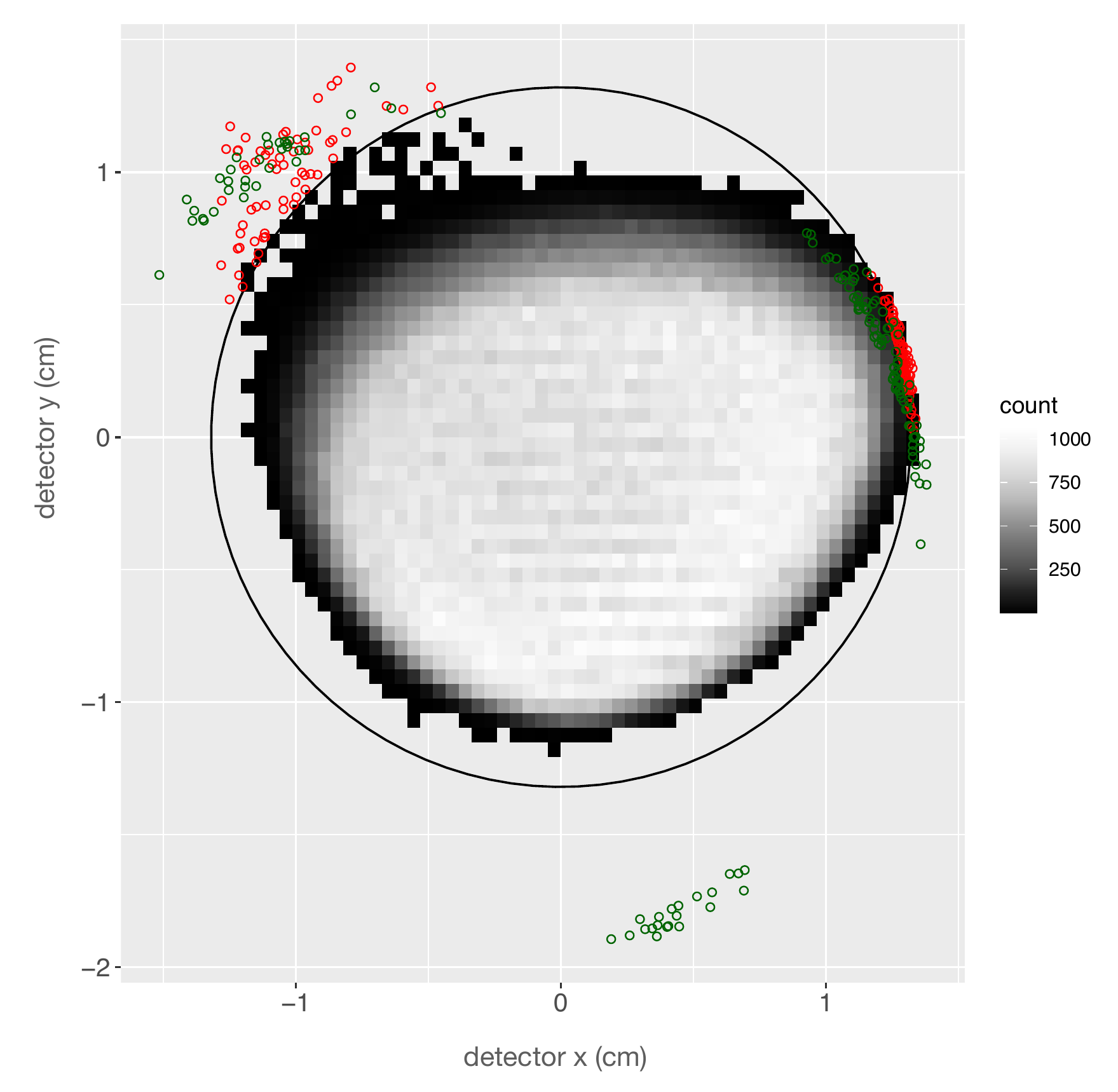}
\caption{\label{F:pFocus} Results from a proton focusing simulation tracking 1 million neutron decay protons from the proton collimator to the detector. Green and red circles are protons striking the focusing ring and detector inactive region, respectively. The thin black circle indicates the active region of the surface barrier detector.}
\end{figure}

We fabricated a set of thin aluminum detector masks that blocked different regions of the detector face. One of these (the ``R4'' mask) blocked a central circle 24.8 mm in diameter, leaving a ring of width 3 mm at the outer edge of the active region exposed to detect protons. Neutron decay data were collected with the various masks installed in 1--2 day runs. The resulting background-subtracted wishbone event rates were compared to the rates found in the simulation using the same mask geometries which enabled us to fix the absolute position of the detector system in space relative to the neutron beam and collimator. The simulation then computed the focusing efficiency. Figure \ref{F:pFocus} shows a simulation of $10^6$ neutron decay protons, out of which 146 struck the focusing ring (green circles) and 154 struck the inactive region of the detector (red circles). No protons missed the detector assembly entirely. The resulting focusing efficiency was 99.97 \%. A 45-hour run with the R4 mask in place produced a wishbone event rate of $(3.8\pm1.9) \times 10^{-3}$ s$^{-1}$, or $(0.33\pm0.17)$ \% of the normal unmasked rate, consistent with the AMaze simulation.
\par
From the simulation of the B$_{\rm up}$ proton assembly the systematic error in the wishbone asymmetry was determined to be $\Delta X / X = -0.0042 \pm 0.0058$, including a 20 \% quadrature uncertainty for the Monte Carlo.
\par
Approximately 0.5 \% of protons incident on the detector are expected to backscatter without producing a countable signal. This occurs at the full kinetic energy 30 keV, where the relative energy difference between the fast and slow groups (about 380 eV)  is small, so the associated systematic error due to proton backscatter is negligible.
\subsection{Wishbone Asymmetry Result, Magnetic Field Up}
\begin{figure}
\centering
\includegraphics[width = 6in]{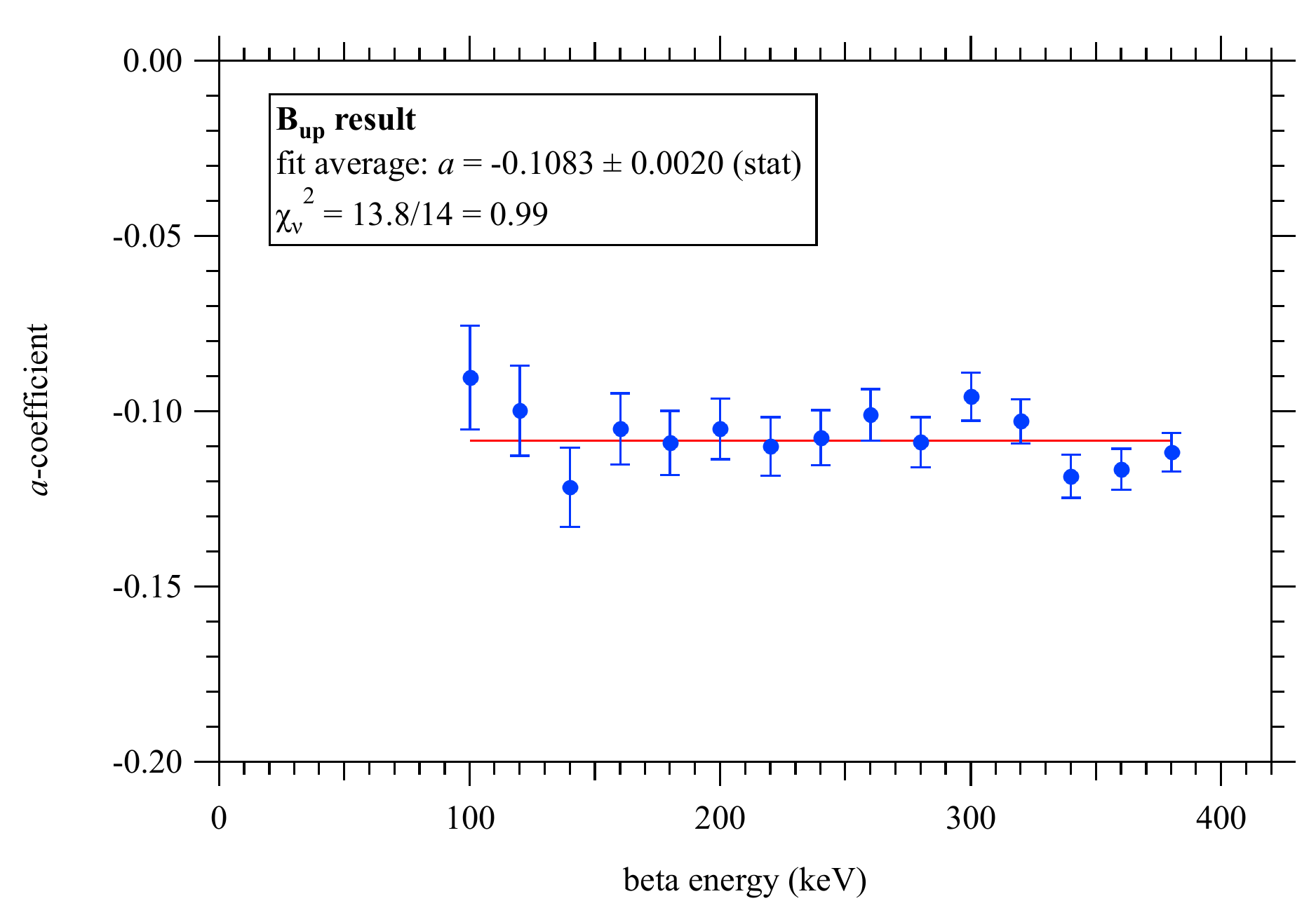}
\caption{\label{F:BupResult} The corrected B$_{\rm up}$ wishbone asymmetry (see figure \ref{F:wbAsymUp}), divided by the geometric function $f_a(E)$ (see figure \ref{F:faE}), giving the measured $a$-coefficient for
each beta energy slice. These were fit to a constant to produce the $a$-coefficient result for the B$_{\rm up}$ data. Error bars are statistical.}
\end{figure}

To produce the corrected wishbone asymmetry, we started with $\left[X(E) - \delta_2(E)\right] / \left[1 + \delta_1(E)\right]$ (see equation \ref{E:aEffective}) and added the systematic corrections described above. This 
can be seen in figure \ref{F:wbAsymUp}. The corrected wishbone asymmetry was then divided by the geometric function $f_a(E)$ to give the measured value of the $a$-coefficient for each energy slice, shown in figure \ref{F:BupResult}. These were then fit to a constant to obtain the overall result
\begin{equation}
a = -0.10834 \pm 0.00197 (\mbox{stat}) \pm 0.00156 (\mbox{sys}) \quad \mbox{(B$_{\rm up}$)}.
\end{equation}

\subsection{Wishbone Asymmetry Analysis, Magnetic Field Down}
\label{SS:asymDn}
\begin{figure}
\centering
\includegraphics[width = 5.5in]{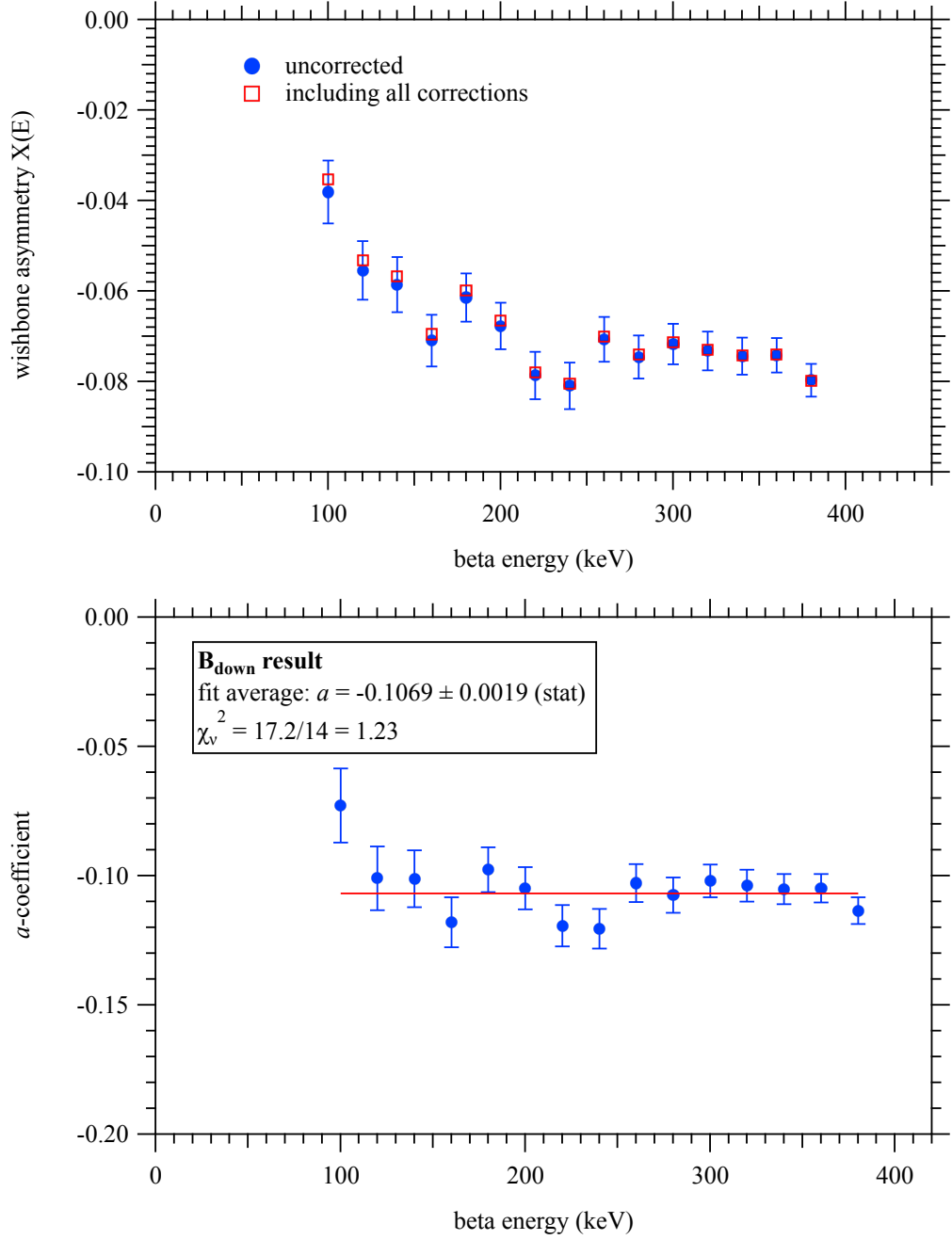}
\caption{\label{F:BdownResult} Top: The wishbone asymmetry $X(E)$ for the combined $B_{\rm down}$ data, uncorrected with statistical error bars, and with all corrections.
Bottom:  The corrected B$_{\rm down}$ wishbone asymmetry, divided by the geometric function $f_a(E)$, giving the measured $a$-coefficient for
each beta energy slice. These were fit to a constant to produce the $a$-coefficient result for the B$_{\rm down}$ data. Error bars are statistical.}
\end{figure}

After finalizing the B$_{\rm up}$ result, we analyzed the B$_{\rm down}$ data in the same way, except that four systematic effects were analyzed independently for B$_{\rm down}$:
\begin{enumerate}
\item {\bf Magnetic field shape: } In the B$_{\rm down}$ field maps the average transverse magnetic field magnitude was 2 $\mu$T, a factor of two larger than
in the B$_{\rm up}$ field maps, so the systematic correction was correspondingly larger, and as before we assign an uncertainty equal to the correction, giving $\Delta X = (-1.7 \pm 1.7)  \times 10^{-4}$.
\item {\bf Absolute magnetic field:} Independent $^3$He NMR measurements were made in the B$_{\rm down}$ configuration with the result B$_{\rm axial}$ = 0.03624(11) T. Because the geometric function $f_a(E)$ was calculated
using B$_{\rm axial}$ = 0.0364 T, a correction of ($0.4\pm0.3$) \% to the wishbone asymmetry was needed.
\item {\bf Proton scattering:} While the effect of proton scattering from the collimator should be the same for B$_{\rm up}$ and B$_{\rm down}$, it was analyzed independently using the method described in section \ref{SSS:pScatter}. The count rate difference in 1-$\mu$s wide regions just before and after the wishbone TOF peak was smaller, $-26 \pm 2272$ counts, leading to a smaller estimate for the correction: $\Delta X = 0\pm0.00034$.
\item {\bf Proton focusing:} In order to accomodate the change in sign of the $E \times B$ force, a separate proton focusing assembly with slightly different geometry was used for the B$_{\rm down}$ run. The systematic error in the wishbone asymmetry was estimated from the $B_{\rm up}$ analysis to be $\Delta X / X = 0 \pm 0.00110$, including a 20 \% quadrature uncertainty for the Monte Carlo.
\end{enumerate}
All other systematic corrections and uncertainties were the same as described in section \ref{SS:Systematics}. The wishbone asymmetry $X(E)$ for the combined $B_{\rm down}$ data, both uncorrected (blue dots) with statistical error bars, and with all corrections (red squares), are shown in the top plot of figure \ref{F:BdownResult}. The bottom plot shows the corrected wishbone asymmetry, divided by the geometric function $f_a(E)$, giving the measured $a$-coefficient for each beta energy slice. These were fit to a constant to produce the overall $a$-coefficient result for B$_{\rm down}$
\begin{equation}
a = -0.10690 \pm 0.00187 (\mbox{stat}) \pm 0.00180 (\mbox{sys}) \quad \mbox{(B$_{\rm down}$)}.
\end{equation}

\section{Result and Discussion}
The difference in the results from the B$_{\rm up}$ and B$_{\rm down}$ runs is
\begin{equation}
a(\mbox{B$_{\rm down}$}) - a(\mbox{B$_{\rm up}$}) = 0.0014 \pm 0.0027 (\mbox{stat}).
\end{equation}
Attributing this difference to a residual neutron polarization gives $P = (5.0 \pm 9.5) \times 10^{-4}$,  consistent with zero, using equation \ref{E:XEpol}. At this point in the analysis we unblinded by revealing the directly measured neutron polarization, $P < 4.0 \times 10^{-4}$ (90 \% C.L.), an upper limit that confirmed the null polarization. The direct neutron polarization measurement on NG-C is described in detail in another publication \cite{Sch20}. We combine the B$_{\rm up}$ and B$_{\rm down}$ results for the aCORN NG-C run
\begin{equation}
a = -0.10758 \pm 0.00136 (\mbox{stat}) \pm 0.00148 (\mbox{sys}) \quad \mbox{(NG-C combined)}.
\end{equation}
The error budget for the combined result is shown in Table \ref{T:Errors}. In producing this table we used the standard deviation of the mean for the independent systematic uncertainties, {\em i.e.} the enumerated list in section \ref{SS:asymDn}.

\par
This result is in good agreement with the result of the aCORN NG-6 run: $a = -0.1090 \pm 0.0030\mbox{(stat)}\pm 0.0028\mbox{(sys)}$ \cite{Dar17}. We may combine them to obtain an overall result from the two
completed aCORN physics runs. To combine these two we first compute the weighted average value of the $a$-coefficient, using statistical uncertainties only. The only systematic correction and uncertainty that was applied equally to both measurements was the effect of electron energy loss in the positive grid of the electrostatic mirror; the others were all evaluated independently. Therefore we remove the grid uncertainty from both, compute the standard deviation of the mean of the two systematics uncertainties, and then add the grid uncertainty back in quadrature. The result is
\begin{equation}
a = -0.10782 \pm 0.00124 (\mbox{stat}) \pm 0.00133 (\mbox{sys}) \quad \mbox{(NG-6 + NG-C combined)},
\end{equation}
or with the statistical and systematic uncertainties combined in quadrature: $a = -0.10782 \pm 0.00181$, for a relative uncertainty of 1.7 \%. Using equation \ref{E:SMtaABD} we can extract a result for $\lambda = G_A/G_V$,
\begin{equation}
\lambda = -1.2796\pm 0.0062 \quad \mbox{(NG-6 + NG-C combined)}.
\end{equation}

\begin{table}
\caption{\label{T:Errors} A summary of systematic corrections and uncertainties for the value of the $a$-coefficient in the combined NG-C result. The third column lists the absolute uncertaintes and the fourth
column is relative to our final result for $|a|$. The combined uncertainty is the quadrature sum of statistical and systematic.}
\centering
\begin{ruledtabular}
\begin{tabular}{lD{.}{.}{5}D{.}{.}{5}D{.}{.}{5}}
systematic & \multicolumn{1}{c}{correction} &  \multicolumn{1}{c}{$\sigma$ uncertainty} & \multicolumn{1}{c}{relative uncertainty}\\\hline
$e$ scattering  & -0.00083 & 0.00083 & 0.0077\\
wishbone asymmetry &  & 0.00064 & 0.0060\\
residual gas &  & 0.00048 & 0.0045\\
proton scattering  & & 0.00038 & 0.0035\\
beta energy calibration  & & 0.00030 & 0.0028\\
electrostatic mirror & 0.00161 & 0.00032 & 0.0030\\
absolute magnetic field & 0.00023 & 0.00023 & 0.0002\\
energy loss in grid & -0.00111 & 0.00022 & 0.0020\\ 
proton collimator alignment & 0.00046 & 0.00020 & 0.0019\\
magnetic field shape & 0.00018 & 0.00011 & 0.0010\\
electrostatic mirror alignment & 0.00025 & 0.00009 & 0.0008\\
neutron beam density & -0.00045 & 0.00009 & 0.0008\\
proton focusing  & 0.00036 & 0.00055 & 0.0051\\\hline
total systematic & 0.00070 & 0.00148 & 0.0137\\
statistical & & 0.00136 & 0.0126\\\hline
combined uncertainty & & 0.00201 & 0.0186\\
\end{tabular}
\end{ruledtabular}
\end{table}
\par
\begin{figure}
\centering
\includegraphics[width = 6in]{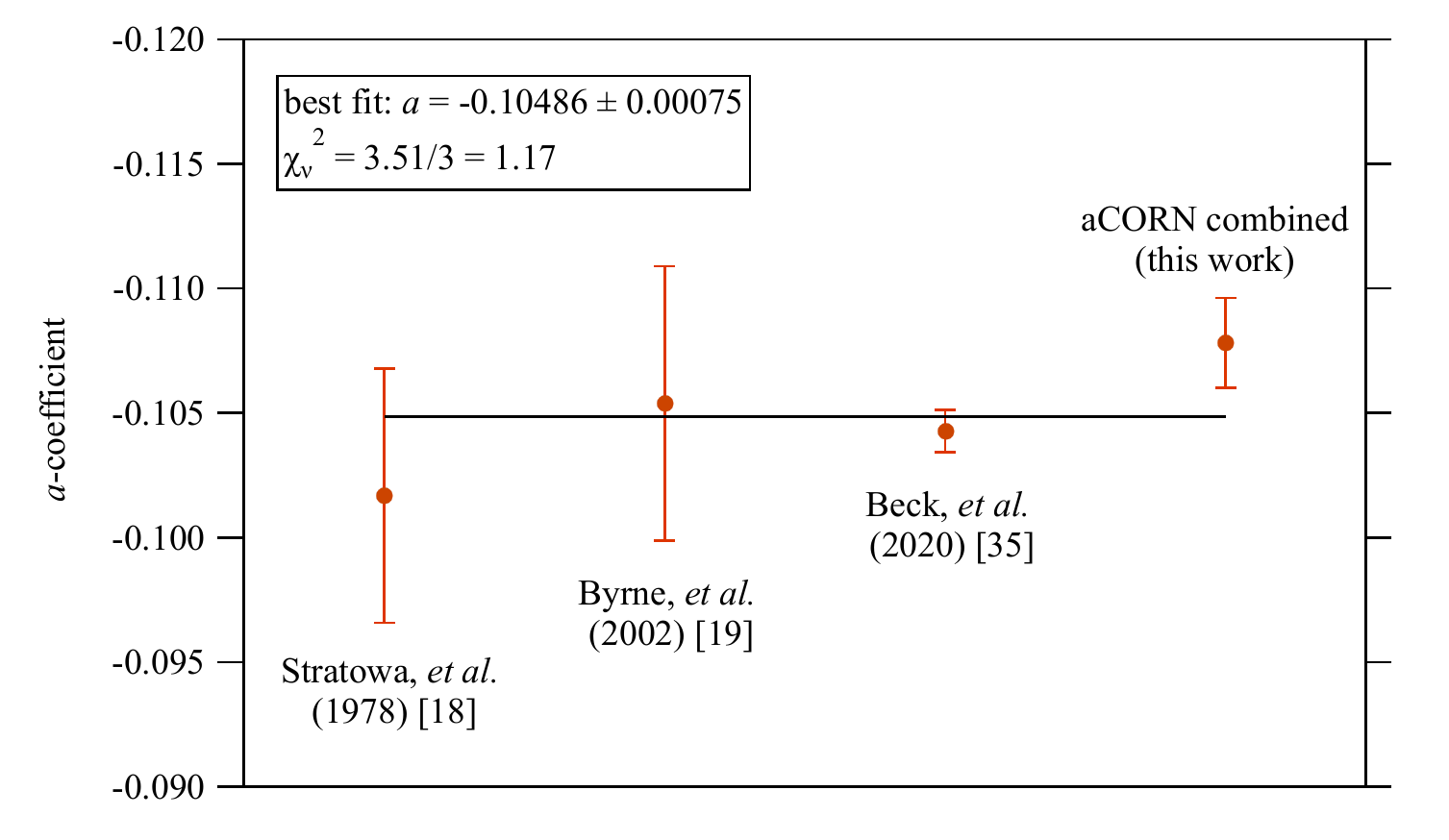}
\caption{\label{F:aSummary} A summary of neutron $a$-coefficient measurements from the past 50 years.}
\end{figure}
\begin{figure}
\centering
\includegraphics[width = 5in]{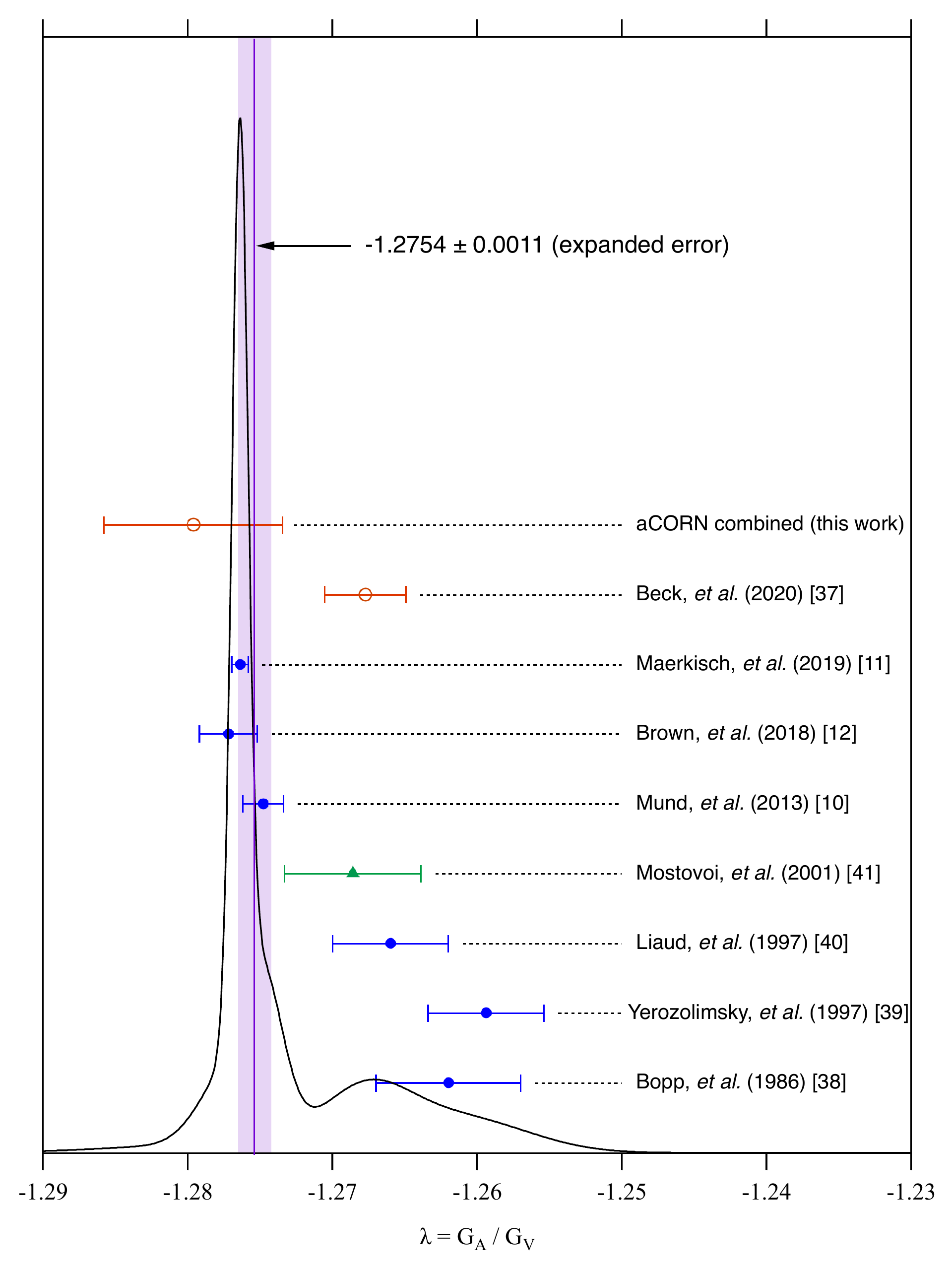}
\caption{\label{F:ideogram} An ideogram of precision determinations of the neutron decay ratio of axial vector to vector coupling ($\lambda$) using the beta asymmetry ($A$-coefficient, blue circles), the electron-antineutrino correlation
($a$-coefficient, red open cirles), and the $A/B$ ratio (green triangle). The distribution features two groups of experimental results and the overall agreement is poor. The weighted average is indicated with the uncertainty expanded by a factor of 2.32.}
\end{figure}

Figure \ref{F:aSummary} shows a summary of four neutron $a$-coefficient measurements from the past 50 years. The 2020 result from the aSPECT experiment \cite{Bec20}, which used an electromagnetic retardation 
spectrometer to measure the proton energy spectrum, is the most precise. The overall agreement of these is good in spite of the slight tension (1.7$\sigma$) between the aSPECT and aCORN results. The weighted  average of these is
\begin{equation}
\label{E:aWorld}
a = -0.10486 \pm 0.00075 \quad \mbox{(world average)}.
\end{equation}
The effects of the new $a$-coefficient results on the world average for $\lambda$ are less satisfactory. Figure \ref{F:ideogram} shows an ideogram, in the style of the Particle Data Group (\cite{PDG20}, p.~16), of precise determinations of 
$\lambda = G_A/G_V$ from the neutron decay beta asymmetry ($A-$coefficient) \cite{Bop86,Yer97,Lia97,Mun13,Bro18,Mar19} and the electron-antineutrino correlation ($a$-coefficient) \cite{Bec20} and this work. Also included is a determination from the ratio of the $A$-coefficient to $B$-coefficient in a combined experiment \cite{Mos01}. The distribution is unfortunately bimodal with poor overall agreement ($\chi_{\nu}^2 = 43.98/8 = 5.37$). The weighted world average is
\begin{equation}
\lambda = -1.2754 \pm 0.0011 \quad \mbox{(world average)}
\end{equation}
with the uncertainty expanded by a factor of $\sqrt{5.37} = 2.32$. 
The aSPECT result adds weight to the more positive number favored by older beta asymmetry experiments. The aCORN result is in better accord with recent beta asymmetry experiments. In particular it is troubling that the most precise results for the $A$- and $a$-coefficients \cite{Mar19,Bec20}, both published within the past two years, disagree by 3 standard deviations. New precision experiments, in particular additional measurements of the neutron $a$-coefficient at the $<$1 \% level, are needed to resolve this. The upcoming Nab experiment \cite{Nab} and a possible future aCORN run at NIST are hoping to achieve such precision.
\par
Finally we can update the values of the Mostovoy parameters (equations \ref{E:F1F2}) using the new world average for the $a$-coefficient (equation \ref{E:aWorld})
\begin{eqnarray}
\label{E:F1F2new}
F_1  & = & 1 + A - B - a = 0.0046 \pm 0.0031 \nonumber\\
F_2  & = & aB - A - A^2 = 0.00244 \pm 0.00081.
\end{eqnarray}
The value of $F_2$ now exceeds zero by $3\sigma$ indicating a strong deviation, for the first time using this test, from the Standard Model prediction. This follows mainly from the disagreement in the value of $\lambda$ between aSPECT \cite{Bec20} and PERKKEO III \cite{Mar19}.

\section{Acknowledgements}
This work was supported by the National Science Foundation, U.S. Department of Energy Office of Science, and NIST (US Department of Commerce). We thank the NCNR for providing the neutron facilities used in this work, and for technical support, especially Eli Baltic, Daniel Ogg, Dan Adler, George Baltic, and the NCNR Research Facilities Operations Group.

\newpage
\newpage
\newpage
\newpage
\newpage
\newpage
\newpage
\newpage
\newpage
\newpage


\begin{thebibliography}{99}
\bibitem{DISCLM} Certain trade names and company products are 
mentioned in the text or identified in illustrations in order to 
adequately specify the experimental procedure and equipment used. In 
no case does such identification imply recommendation or endorsement 
by the National Institute of Standards and Technology, nor does it imply 
that the products are necessarily the best available for the purpose.
\end{thebibliography}
\end{document}